# An AI-powered Tool for Central Bank Business Liaisons: Quantitative Indicators and On-demand Insights from Firms

Nicholas Gray, Finn Lattimore, Kate McLoughlin and Callan Windsor

Research Discussion Paper
2025–[##]

June 2025

Economic Group
Reserve Bank of Australia

We would like to thank Markus Schwedeler, James Holloway and Claude Lopez for their helpful feedback and suggestions as well as seminar participants at the RBA/APRA Business Coding Knowledge Sharing Sessions and internal RBA seminars. There are also several people to thank for their contributions to the development of this tool. Max Zang, for his work on numerical extractions; Oliver Cutbill, for his work on the uncertainty index; Maureen Rocas, for helping to establish initial data pipelines; Joel Fernando, for his work on topic exposures; and Madison Terrel, for her patience and expertise helping us understand and source data to replicate current nowcasting processes at the RBA. We are also grateful to the entire Regional Industry and Analysis team (past and present) for their ongoing dedication to help us improve the tool and to all firms who participate in the RBA's liaison program. These are the views of the authors and not the Reserve Bank of Australia. Any remaining errors are our own.

Authors: windsorc at domain rba.gov.au

Media Office: rbainfo@rba.gov.au

Please cite this work as: **Gray N, F Lattimore, K McLoughlin and C Windsor (forthcoming)**, 'An AI-powered Tool for Central Bank Business Liaisons: Quantitative Indicators and On-demand Insights from Firms', RBA Research Discussion Paper No 2025-xx.

**Abstract**

In a world of increasing policy uncertainty, central banks are relying more on soft information sources to complement traditional economic statistics and model-based forecasts. One valuable source of soft information comes from intelligence gathered through central bank liaison programs — structured programs in which central bank staff regularly talk with firms to gather insights. This paper introduces a new text analytics and retrieval tool that efficiently processes, organises, and analyses liaison intelligence gathered from firms using modern natural language processing techniques. The textual dataset spans 25 years, integrates new information as soon as it becomes available, and covers a wide range of business sizes and industries.

The tool uses both traditional text analysis techniques and powerful language models to provide analysts and researchers with three key capabilities: (1) quickly querying the entire history of business liaison meeting notes; (2) zooming in on particular topics to examine their frequency (topic exposure) and analysing the associated tone and uncertainty of the discussion; and (3) extracting precise numerical values from the text, such as firms' reported figures for wages and prices growth.

We demonstrate how these capabilities are useful for assessing economic conditions by generating text-based indicators of wages growth and incorporating them into a nowcasting model. We find that adding these text-based features to current best-in-class predictive models, combined with the use of machine learning methods designed to handle many predictors, significantly improves the performance of nowcasts for wages growth. Predictive gains are driven by a small number of features, indicating a sparse signal in contrast to other predictive problems in macroeconomics, where the signal is typically dense.

JEL Classification Numbers: E5, E6, J3, C5, C8, D2

Keywords: central banking, macroeconomic policy, wages and labour costs, machine learning, econometric modeling, information retrieval systems, firm behaviour

# Table of Contents

## 1.    Introduction

Central bank business liaison programs play an important role in shaping monetary policy decisions by providing timely, on-the-ground insights on economic conditions. Since 2001, staff from the Reserve Bank of Australia (RBA) have conducted around 22,000 interviews with a broadly representative sample of firms, industry bodies, government agencies and community organisations (hereafter referred to as 'firms') from across the country under the banner of the RBA's liaison program.

This formal program of economic intelligence gathering has been a useful complement to published economics and finance statistics and information gleaned from econometric models in informing the RBA's assessment of economic conditions. For example, on average, since the pandemic there have been around 550 references to liaison in the RBA's policy-related publications. Figure 1 shows that references to liaison in the RBA's *Statement on Monetary Policy* have grown over time, reflecting that the RBA has increased its reliance on liaison over the past decade. The real-time availability of liaison information, as well as the richness of the qualitative insights it provides, are particularly useful for navigating periods of heightened uncertainty.

**Figure 1: Mentions of Liaison in the *Statement on Monetary Policy***
Share of total words

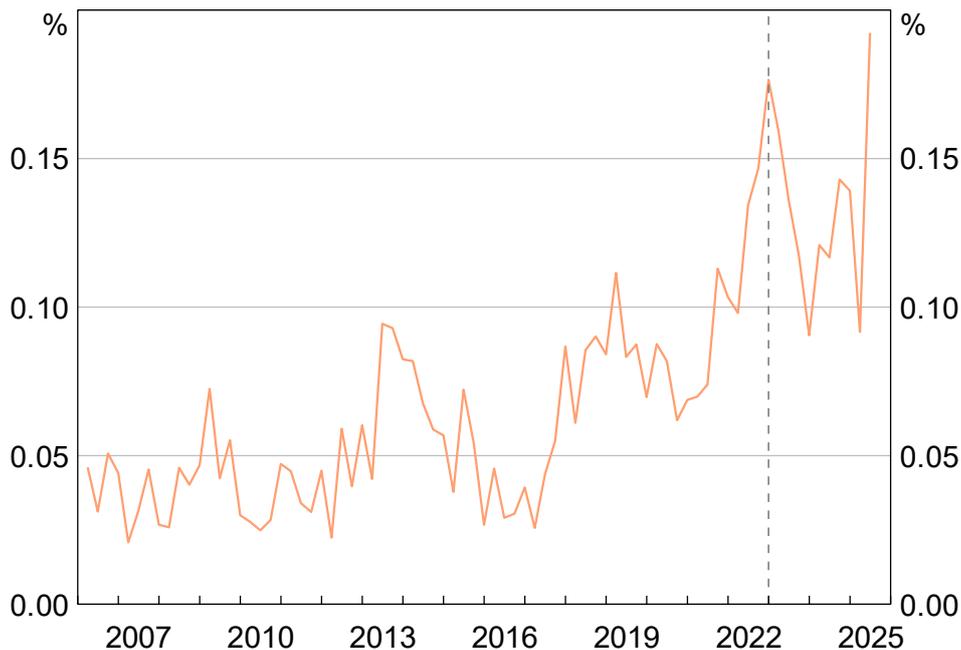

Note:    Ratio of the number of mentions of liaison in the RBA's *Statement on Monetary Policy* to the total number of words. Vertical dashed line indicates the introduction of a dedicated "Insights from Liaison" Box.

Source:    RBA

As policy makers face into more uncertain and complex macroeconomic conditions (see, for example, Lagarde (2023) and Wolf (2023)) it is critical that they can extract signals efficiently and effectively from textual sources of information from firms, such as earnings reports, earnings calls, and information from interviews and focus groups. Analysing these sources of information vastly expands the available information from conventional quantitative data alone, containing information on 'why' and 'how' firms are responding to unexpected events, decision-makers' expectations, perceived risks and future opportunities (Hassan *et al* 2024). However, this first-hand information is typically subjective, contextual and is not designed and collected for statistical purposes. As such, it can be difficult to quickly systematically synthesise and communicate the high frequency signals offered by such data in a way that is useful for informing decisions. Reliance on narrative information also exposes decision making to the risk of unconscious cognitive and behavioural biases.[1] Fortunately, recent developments in natural language processing (NLP) offer the prospect of capturing the nuance and richness in this type of information in an efficient and more systemic manner. This increases both the breadth and depth of the analysis and can reduce exposure to unconscious cognitive and behavioural biases.[2]

In this paper, we introduce the RBA's text analytics and retrieval tool for quantitative and qualitative analysis of liaison messages. The new tool was created in late 2022 and has begun to be used to support economic analysis. It uses modern techniques in NLP to improve analysts' abilities to quickly search and extract signals from liaison text about economic conditions. In particular, the tool introduces three new capabilities: (1) the ability to quickly query the entire history of liaison meeting notes spanning around 22,000 liaison meetings over the past 23 years; (2) the ability to zoom in on particular discussion topics (for example, "supply chains") to examine the level of interest in the topic over time and the associated tone and uncertainty of the discussion; and (3) the ability to extract precise numerical values from the text, such as firms' self-reported growth in wages and final prices. The information underpinning the tool is automatically updated daily, drawing in text data from new liaisons once the meeting notes are finalised. Almost instantaneous insights, at the firm-level as well as aggregated, can then be accessed using a simple no-code application.[3]

The first aim of this paper is to offer a detailed overview of the technical architecture underpinning our text analytics and retrieval tool for central bank liaison information as well as disseminate the techniques we use to enrich and extract signals from the associated text. In doing so, we hope to contribute to the development and deployment of similar tools across the central banking community. To this end, replicable code for the back and front-end of our tool as well as for our nowcasting exercise is publicly available on GitHub (*forthcoming)*.

We also contribute to the literature on textual analysis and information retrieval systems for macroeconomics. To our knowledge, this paper is the first to apply modern techniques in NLP to develop an analytics and information retrieval tool for confidential and (near) real-time intelligence gathered through central bank business liaison programs. In doing so, we build on a growing body of work by scholars, central banks, think tanks and international organisations using techniques in NLP to draw insights from other text-based sources of corporate information, such as earnings calls, to address macroeconomic questions related to policy as well as for their economic surveillance activities. Recent examples include the use of quarterly earnings calls to analyse the effect of trade exposures on bank lending (Correa *et al* 2023); firms' price-setting behaviour (Windsor and Zang 2023); firm-level climate exposures (Sautner *et al* 2023); supply and demand imbalances (Gosselin and Taskin 2023); cyber risk exposures (Jamilov, Rey and Tahoun 2021); political risk exposures (Hassan *et al* 2019); sources of country risk (Hassan *et al* 2021); and the diffusion of disruptive technologies (Bloom *et al* 2021). In its World Economic Outlook, the International Monetary Fund has used text from earnings calls to monitor firm-level inflation expectations (Albrizio *et al* 2023), while the European Central Bank and RBA are using earnings calls to monitor various aspects of firm sentiment (see RBA 2022a and Andersson *et al* 2023). NLP techniques have also been applied to the Federal Reserve's Beige book, which is a summary of commentary on economic conditions from a wide range of business and community leaders across the United States, gathered as part of their own liaison and business survey program, released eight times a year. The Beige Book has been shown to contain useful information about inflation trends (Gascon and Werner 2022) and economic sentiment (Filippou *et al* 2024).

---

1    These include confirmation bias, conservatism bias, interview bias, measurement bias, near-term bias, observer-expectancy effects, response bias, sampling bias and serial position effects. While these biases can be pervasive and difficult to detect, the RBA's liaison team has adopted several strategies to minimise them in the liaison process and reduce their influence on qualitative and quantitative research and analysis.

2    This notwithstanding, while NLP techniques can mitigate some biases, they are not immune to introducing or perpetuating other biases, for example related to language and context.

3    Meetings conducted by the RBA's liaison staff are undertaken on a confidential basis. This aids in firms being comfortable to share figures and firm-level insights openly with the RBA, as does the trust built over years of engagement with participants. Before being shared more widely across relevant areas of the RBA, information from the liaison program is aggregated, de-identified and summarised. Reflecting the RBA's confidentiality commitments, this information is shared at an industry or economy-wide level. Similarly, only the RBA's liaison staff, data scientists and select technology support workers have access to the tool.

Finally, we make an economic contribution. We explore the benefits of integrating our liaison-based textual indicators into machine learning predictive models, with an application to nowcasting growth in Australia's official wage price index (WPI). Our approach is compared to the current best-in-class baseline Phillips Curve model being used at the RBA. Incorporating liaison-based indicators significantly enhances nowcasting performance for wages growth, increasing accuracy by around 20 per cent over the full sample. In the case of assessing labour market outcomes at least, this validates the ongoing use of liaison to inform the Bank's assessments of economic conditions. Interestingly, the predictive gains appear to be driven by a small number of variables – that is, the signal is sparse. This is an important caveat to the lack of empirical support that has emerged for sparse models in a variety of other predictive applications in macroeconomics (Giannone, Lenza and Primiceri 2021).

The rest of this paper is organised as follows. Section 2 provides background to the RBA's economic liaison program; Section 3 outlines the technical solution architecture underpinning our information tool; Section 4 presents and discusses the new surveillance and analytical capabilities enabled by the tool; Section 5 explores how these capabilities can be used to inform predictive problems by presenting a machine learning exercise based on nowcasting wages growth; and Section 6 concludes.

## 2.    RBA's Liaison Program

In 2001, the RBA established its liaison program – a formal program of economic intelligence gathering. The program involves Bank staff verbally interviewing contacts who typically hold executive roles in organisations.  Interviews are conducted on a continuous basis throughout the year; this contrasts with some business surveys that are conducted at static points, such as the end of the month or the quarter.

On average, around 75 contacts are spoken to a month across the nation and on average around 900 meetings are conducted per year. While most are interviewed for about an hour once per year, 10–15 per cent of contacts are spoken to at shorter intervals. Over a typical year, just over 650 individual contacts are interviewed. The RBA has conducted around 22,000 interviews as part of this program over the past 25 years. Through these interviews, the RBA collects four kinds of economic information.

1. **Basic metadata**: including the firm's name, the broad industry group the firms' activities fall within, the most relevant four-digit Standard Industrial Classification or SIC code (a more detailed industry classifications than industry groupings), headcount, and geographic location of the Head Office or branch office where relevant.

2. **Qualitative views**: prose text of firms' answers to economic questions posed by Bank staff, captured as 'liaison summaries'. Most questions posed over the life of the program have been consistent over time. That is, most participants have been asked at each liaison interview the following 'core' questions: how demand for your goods or services, your investment, headcount, non-labour costs, wages and prices changed over the past year, and how are they expected to change over the coming year? The text captured often includes firms' explanations of 'why' these economic conditions have changed (or not) and what reasons underpin their expectations for economic conditions over the year ahead. The prose captured also includes firms' answers to occasional topical questions.

3. **Quantitative economic outcomes**: many firms also offer numerical percentage firm-level outcomes and expectations in responding to some of the above core questions (e.g. wages and prices growth for a firm).

4. **Staff scores**: Based on the responses provided in the interview and recorded in prose, Bank staff assign a quantitative indicator of the extent of change in the key economic variables being discussed. Specifically, staff assign scores on an ordinal scale from −5 to +5 based on the conditions reported by liaison contacts compared with one year ago and a score for their expectations for the year ahead where that information is provided. If the level of a variable was unchanged, the score would be zero. An extreme rise would be assigned a score of +5 and an extreme fall assigned a score of −5. While judgement is required for the scores in between, a score of +2 would generally be considered a 'normal' or 'average' increase from the previous year.

These records are not exact transcripts of the meeting but are detailed records, for example on average around 1,200 words per meeting have been recorded over the past five years. Records are completed on the basis of what was said from the perspective of the firm, but in recording the comments they are organised thematically and terminology is aligned where appropriate to be consistent with Australian statistical concepts.

The interviews and as such the information recorded tends to be from medium- to larger-sized firms that are broadly representative of the industry and geographic structure of the economy. The liaison program has a pool of around 800 active contacts. Three-quarters of these are businesses, though an important part of the program is liaison with industry associations, government agencies and community associations that together provide a fuller picture of economic conditions. The known gaps in the sample, for example direct exposure to small firms (as opposed to the indirect exposure through industry associations) is a limitation of the dataset.[4]

---

4   For further discussion of the history and nature of the RBA's Economic Liaison Program see: Dwyer et al (2022) 'The Reserve Bank's Liaison Program Turns 21'.

## 3.    Solution Architecture

The solution architecture underpinning our new liaison-based analytics information retrieval tool was designed in close (and ongoing) collaboration between data scientists and economists from the RBA's liaison team. At a high level, staff from the liaison team wanted a tool to help them in three broad areas. First, to be able to answer additional ad-hoc requests from senior decision makers more comprehensively, such as, *"what have we been hearing lately about topic X in industry Y"*? Second, the ability to make fuller use of the complete history of liaison information to place information within a historical context and to synthesise available information more efficiently. Third, to improve the ability to quickly extract signals from liaison text and communicate them more effectively. In this section, we outline the components of the solution and how they come together to achieve these desired objectives. We begin by outlining how the text of a liaison meeting is generated (the data generating process) before detailing our text pre-processing and enrichment steps.

### 3.1    The data generating and enrichment process

Over the life of the liaison program, details of each of the RBA's 22,000 liaisons have been systematically recorded in two information management systems (Figure 2):

- The first is the Client Relationship Management (CRM) database. When a liaison has been conducted, the occurrence of the meeting is recorded in the CRM. This record includes information such as the contact date, metadata about the business (name, industry, etc.), attendees and any staff scores associated with the meeting as discussed in Section 2.

- The second is the RBA's File Management System (FMS), which is where staff save a confidential written summary of each liaison meeting. After drafting the summary according to a pre-defined template, the detailed summaries are reviewed by an editor, who also attended the liaison meeting, and added to the FMS. About one-third of the summaries are completed within two days of the liaison meeting (with this share increasing in the lead up to Monetary Policy board meetings) and most are completed within a week. A unique identifier is generated when a liaison is saved in the FMS and is also added to the CRM database.

**Figure 2: Solution Architecture**

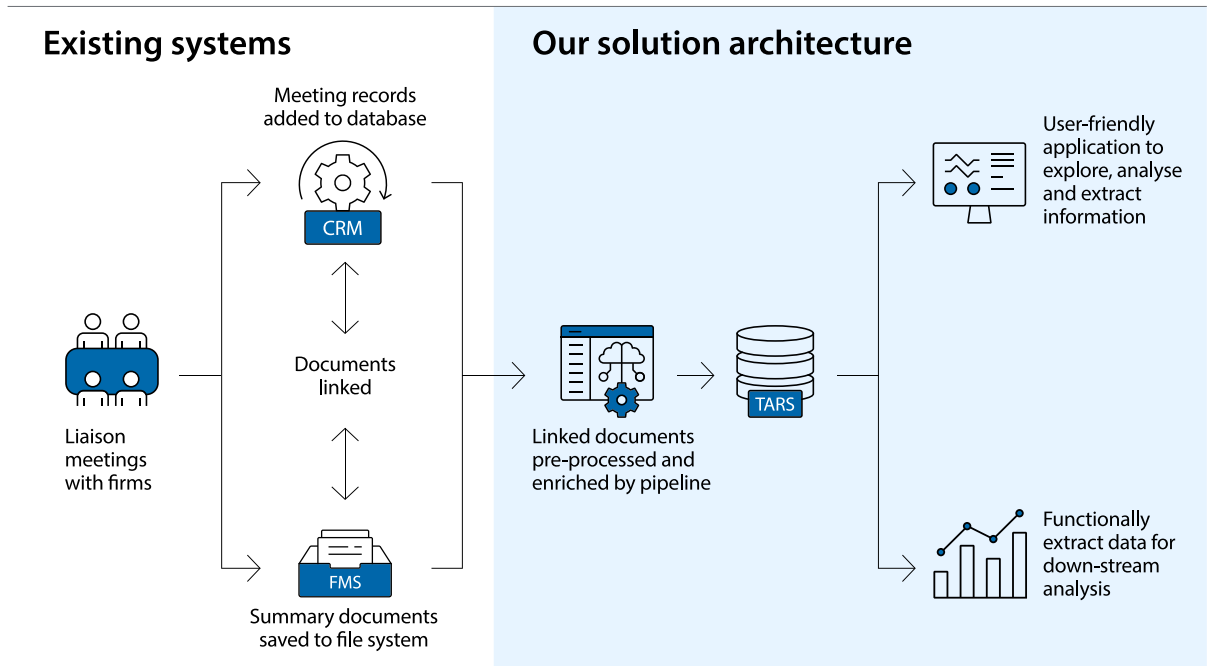

These unique liaison identifiers provide a critical link between these two information management systems, which together form the backbone of our business intelligence text analytics and retrieval system (TARS). To build the TARS, liaison summaries are extracted and cleaned using a text-preprocessing and enrichment pipeline, which does the following:

1. First, liaison summaries are identified and programmatically extracted from the FMS by looking up their unique identifiers as recorded in the CRM.

2. Next, all text summaries are converted to an Extensible Markup Language (.xml) format. This format is useful for several reasons. It is interoperable, meaning the text can be easily shared and processed across different systems and platforms. It also encodes styles inherited from the original text document (e.g. .doc or .docx) such as headings, paragraphs, and other formatting elements. We use these encoded text styles in combination with a set of heuristics to tag each component of each liaison summary as either a "HEADING", "BODY", "TABLE" or "UNKNOWN".[5] "BODY" text chunks include bulleted dot points and paragraphs, both of which will be referred to as paragraphs going forward.

3. Text identified as paragraphs are then enriched with NLP-generated tags. These tags include information about the topic being discussed in the paragraph (e.g. "wages") and the associated tone of the discussion. We also use NLP techniques to extract precise numerical information from the paragraph, such as expected growth in wages or prices over the year ahead. Further details of these NLP methods are outlined in the next section.

---

5   These heuristics were developed in close consultation with subject-matter experts from the business liaison team with intimate knowledge about how the liaison summaries are written.

4. Finally, each enriched paragraph and its associated tags are uploaded into a relational database indexed at a paragraph-level. In addition to the NLP generated tags, every paragraph is also tagged with all the meta-data about the liaison meeting as recorded in the CRM.

The initial synthesis of the large corpus of liaison text into a searchable database required an intensive extraction of historical records from the FMS. Greatly assisting in this process was the stable and centralised record keeping processes that have been consistently followed by the liaison team over time, greatly reducing the complexity of recalling documents up to 23 years old.[6]

## 3.2    Extracting information

Our final TARS database provides the ability for liaison officers to retrieve information quickly and efficiently from the full corpus of liaison intelligence. Insights from these data can be retrieved through a user-friendly application that does not require any coding skills. This makes our TARS accessible to the whole liaison team. In addition to the front-end application, the data can be extracted using several coding languages for more bespoke analytics and requests. The result is that the full corpus of liaison text can be filtered, analysed, and retrieved within seconds.

To keep the TARS database up to date, liaison records are continually added and updated in the CRM and the FMS, which are picked up by a nightly update of the TARS database. This ensures the most up to date information about firms is always available for extraction from the tool.

## 4.    New Capabilities

## 4.1    Quick searches

The first and simplest feature introduced by our new liaison-based TARS is the ability to efficiently filter the full history of liaison text using combinations of keywords and other metadata, such as a firm's geographic location and industry. After filtering, staff can subsequently extract insights for downstream policy analysis and produce on-demand briefings for executive staff.

Previously, the CRM captured a time series of ordinal staff scores and quantitative wage outcomes reported by firms. Staff kept also manual logs of key messages by theme. However, the process of manually retrieving liaison summaries on a broader range of topics and narrowing in on the relevant sections of the text were time-consuming tasks. This made it difficult to respond to as many ad-hoc requests for information as might be useful. Further, doing so in a timely way was in many cases not possible and it was not easy to place information within its historical context, which meant that detailed historical comparisons of liaison information were infrequently completed. These difficulties in systematically searching over the full history of liaison information also exposed staff to potential behavioural biases.

The relational structure of the TARS database means that staff are now able to simultaneously apply many different combinations of paragraph and firm-level filters for highly specific searches of the full history of liaison paragraphs. For example, in Figure 3, the application has been used to show only paragraphs for construction firms in the state of New South Wales that mention either cost, costs or expenses.

---

6   However, as with any legacy data entry system there were cases where the identifier for the document was incorrect or incomplete. Most cases of non-matches occurred in the first few years of the program. Using fuzzy matching of the file identifier, contact name and date to the full list of documents in relevant folders of the FMS increased the matching rate. This notwithstanding, caution should be adopted when using information extracted from the tool prior to 2008.

**Figure 3: A Stylised Example of the Search Interface**

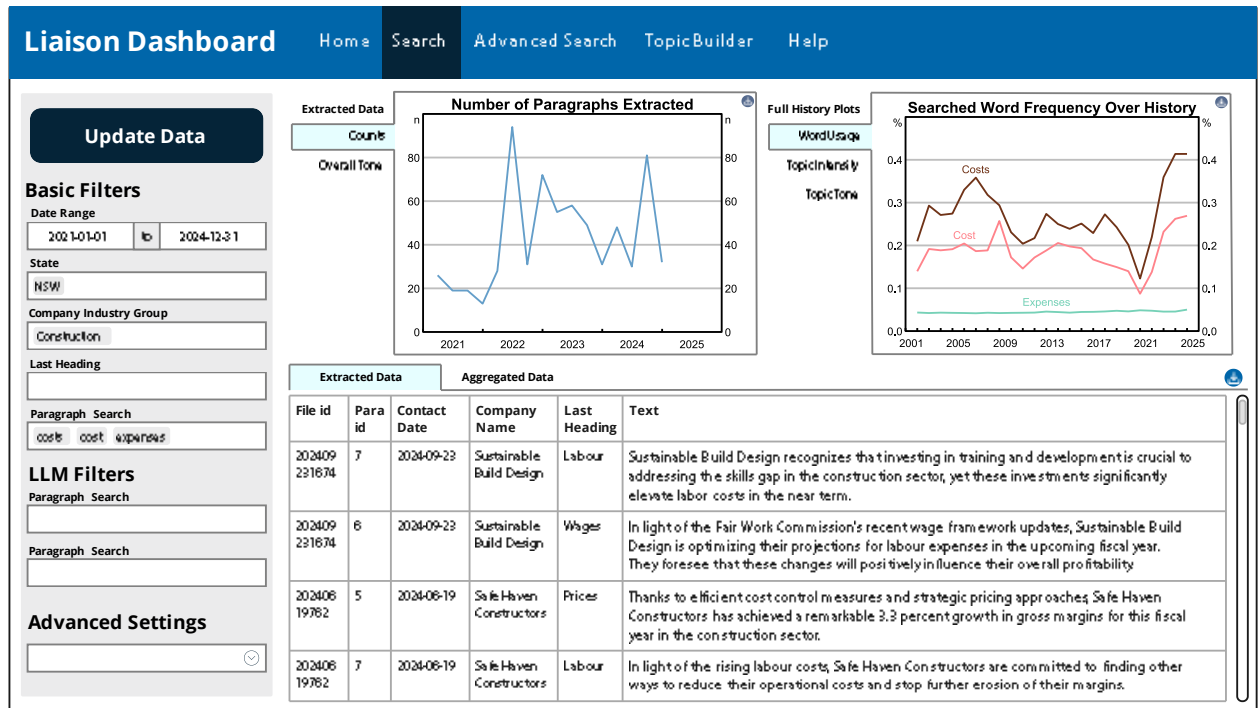

Note: This purely illustrative example shows mock firms and liaison data generated by ChatGPT because of the confidential nature of the information collected and is intended to be illustrative only.

Source: RBA

Advanced filtering can also be used to quickly examine term-frequency indices over the full history of liaison text. For instance, Figure 4 shows a count of the key terms *supply, shipping* and *delays* aggregated to a quarterly level to measure the associated term frequency over time in the liaison dataset.

**Figure 4: Term Frequencies – Selected Supply Chain References**
Share of total words, quarterly

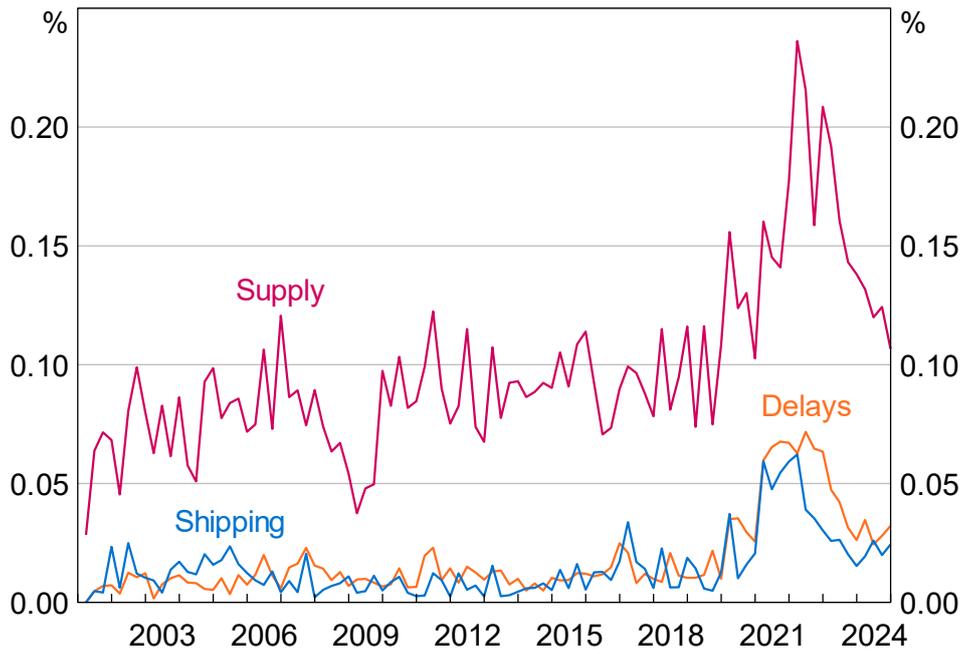

Source: RBA

## 4.2    Topic and tone

The next capability introduced by our liaison-based TARS is filtering the historical liaison summaries by the macroeconomic and financial themes that pervade them. That is, rather than retrieving paragraphs through keyword and meta-data filters alone, users can filter based on the themes that run through the text.

This capability confers two additional benefits. First, filtering the text by broad topics related to macroeconomics and finance allows for more general discovery of information, which is particularly well suited for broader topics of interest. For example, a user interested in firms' costs could filter for the topics of "non-labour costs" and/or "labour costs" to reveal all the possible ways costs may have increased or decreased for firms, both through time as well as between firms at a point in time. By manually parsing the associated text snippets a user could then drill into specific details about the various factors that have affected firms' costs over the cycle.

Second, in addition to filtering by topic and then manually examining the text, statistical indicators can be constructed that measure, in a systematic way, the level of interest in the topic as well as the tone of the associated discussion. The fundamental idea behind these widely used text-as-data statistical measures (see Blei (2012)) is that the amount of time interviewees spend talking about a specific topic indicates its significance and that the tone of the discussion is informative and can be captured by examining the semantics of the text – such as the lexical choice of words, modifiers, polarity, negation and use of intensifiers or diminishers.[7]

Specifically, we can define the level of interest in a topic by counting the number of text snippets (be it sentences or paragraphs) dedicated to it. For a given period $t$, we have liaisons $i = 1...N_t$. The topic exposure for liaison $i$ is:

$$\text{TopicExposure}_i = \frac{\text{Count of snippets on a given topic}_i}{\text{Total number of snippets}_i} \qquad (1)$$

Average topic exposure for all liaisons within period $t$ (including multiple liaisons by the same firm, if applicable) can be computed by averaging this ratio over the $N_t$ liaisons that occurred during period $t$, treating each liaison's exposure equally regardless of the length of the summary:[8]

$$\text{Average TopicExposure}_t = \frac{1}{N_t} \sum_{i=1}^{N_t} \frac{\text{Count of snippets on a given topic}_{i,t}}{\text{Total number of snippets}_{i,t}} \qquad (1.1)$$

Next, the tool allows staff to measure the tone of discussions about a given topic, giving rise to our second NLP-based aggregate indicator:

$$\text{Average TopicTone}_t = \frac{1}{N_t} \sum_{n=1}^{N_t} \text{Average tone of snippets on a given topic}_{i,t} \qquad (2)$$

Equation (2) calculates the average tone of the discussions about a particular topic by averaging the tone of each single liaison and then averaging these over the $N_t$ liaisons that occurred during period $t$.[9]

The procedures we use to classify the topic of each text snippet and measure the tone of the associated discussion are discussed next.

### 4.2.1   Language Models

Our first approach to classify the topics discussed in each liaison paragraph and measure its tone is to use transformer-based language models (LMs).

Starting with topic classifications, for every liaison paragraph – including new paragraphs as they become available – we use a transformer-based LM to determine the probability the paragraph is about each of our 14 topics of interest. Each topic probability score is between 0 and 1 and is independent of the other topic scores, meaning a paragraph can have a high likelihood of being about multiple topics at the same time. In practice, this is done by enabling the model's multi-label option, such that it performs 14 separate passes over the corpus to assess the probability of each topic independently. In the liaison summaries each paragraph of text is often about multiple topics and we are more interested in identifying those topics rather than clearly distinguishing between predominant topics.

---

7    The focus of liaison meetings is primarily driven by the core questions that have been stable over time, with a portion of meetings typically used to answer each question. Interviewers can steer interviewees with follow-up questions, but by-and-large, the length of time spent on a topic is determined by a contact's interest and willingness to share information on a topic with the vast majority of each interview comprised of the interviewee talking.

8    Alternatively, overall topic exposure can be calculated by summing the counts of relevant text snippets on the given topic across all firms at time $t$ and dividing by the total number of snippets across all firms at time $t$, providing a collective measure of topic exposure, where firms with longer liaisons have more effect on the overall exposure.

9    In Equation (2), firms are not weighted according to how many text snippets they dedicate to discussing the topic of interest. To account for this, overall topic-specific tone can be calculated by taking an average of the tone of all relevant text snippets in time period $t$, with firms that discussed the relevant topic more having more effect on the overall exposure.

The chosen topics of interest include:

- "demand", "sales", "investment or capex", "property or housing", "employment", "wages", "prices", "margins", "costs", "labour costs", "non-labour costs", "supply chains", "financing conditions", and "climate change".

These topics were selected in close collaboration with subject-matter experts from the liaison program and broadly reflect the core questions used in the program. Categories can be expanded to include new topics as needed.[10] The stylised example shown in Figure 5 illustrates how the model assigns probability scores to paragraphs of text, with the LM in this instance indicating there is a high probability the paragraph is about "employment", "wages" and "labour costs". The specific transformer-based LM used for topic classification is BART-large-MNLI.[11] This model can classify text into user-defined categories without prior training on those specific categories.[12] The model's training objective is to minimise the statistical loss associated with re-constructing corrupted sentences into their original form, learning the dependencies and context of natural language along the way (Lewis *et al* 2019).

---

10  Any arbitrary topic label can be added to the list. With the computational resources available to us now (T4 GPU) classifying the current history of paragraphs into a new topic takes around 2.5 hours.

11  The Huggingface model card is here: link.

12  BART-large, a pre-trained model based on Bidirectional and AutoRegressive Transformers, learns language semantics from large datasets such as the entire Wikipedia and BookCorpus. Bidirectional and AutoRegressive Transformers are advanced types of neural networks used in NLP. "Bidirectional" means the model reads text in both directions (left-to-right and right-to-left) to understand context better. "AutoRegressive" means the model predicts the next word in a sequence based on the previous words. Together, these techniques help the model understand human-like text. The MNLI part, which stands for Multi-Genre Natural Language Inference, fine-tunes BART using a dataset of sentence pairs labelled as entailment or contradiction. This fine-tuning enables BART to function effectively as a zero-shot text classifier. Despite the name, BART-large is now considered a small language model – with only around 400 million parameters. We opted for a small LM to ensure fast processing across the entire data history, staying within the compute available in our secure environment. As we gain access to improved computational capacity, future work may examine whether larger models can improve performance.

## Figure 5: An Example Classification from the Language Model

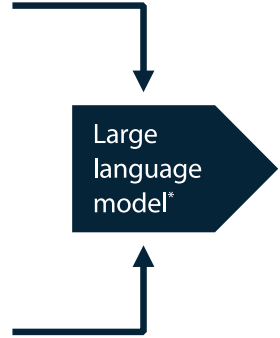

To filter paragraphs based on the output from the topic classification model we need to map from the probabilistic scores to a true/false variable that indicates if a given paragraph is about a given topic. We do this with a threshold: a topic label is assigned to a paragraph if the probability for that label is greater than a pre-defined threshold:[13]

$$\text{Filter} = \begin{cases} True & \text{if } Prob(topic_p) > \text{threshold} \\ False & \text{otherwise} \end{cases} \tag{3}$$

We can also quantify the level of interest in each topic over time. For liaison $i$ in time $t$ the count of paragraphs $p$ about a given topic is given by:

$$\sum_{p=1}^{P_{i,t}} \mathbf{1}\{Prob(\text{topic}_p) > \text{threshold}\} \tag{4}$$

where the indicator function 1{.} is equal to +1 if the topic probability is greater than some threshold. This expression is the numerator in Equation (1) above when it is estimated using LMs.

Next, we use a transformer-based LM called FinBERT (Yang, UY and Huang 2020) to analyse the tone of each paragraph. FinBERT is fine-tuned to assess the sentiment of financial text. It was trained on a large dataset of financial news, enabling it to capture the nuances of financial language more accurately. FinBERT outputs a sentiment score between -1 (negative) and +1 (positive) for each paragraph.[14] The combination of topic classification and sentiment labels for each paragraph provides a comprehensive view of the interest in, and tone of, different economic concepts. Users can select paragraphs based on the topic classification (Equation (3)) and then calculate the average tone of these paragraphs by averaging over the sentiment scores estimated by the LM.

---

13  The application allows users the flexibility to choose the threshold, but we recommend setting the threshold high enough to minimise the incidence of false positives.

14  FinBERT has a reported accuracy of 84 per cent on the Financial PhraseBank dataset classification task, beating available benchmarks at the time of publication.

### 4.2.2 Keywords

Our second approach to classify the topics in each liaison paragraph and measure its tone is to use lists of relevant keywords.

Starting with thematic categorisations, users can construct their own topics by building up lists of keywords or phrases related to that topic. One way to construct these lists of keywords is to consult with analysts that have deep domain expertise in the topic of interest. For example, this was the approach taken in Windsor and Zang (2023) when constructing lists of keywords related to input costs, demand and final prices. These lists of keywords can be passed into our tool to filter paragraphs.

As a complement to this approach, our tool helps users augment their lists of topic-specific keywords using machine learning. Specifically, it uses Word2Vec (Mikolov *et al* 2013) to automatically suggest words that are semantically similar to those in a pre-existing list of keywords or phrases (see Figure 6, for example) to help analysts build up the most representative list of words for a topic. In this way, users can iteratively use the pre-trained Word2Vec model to populate and extend a list of keywords by first specifying an initial set of seed words related to the topic; second, selecting which suggested words (based on semantic similarity) to incorporate; and third, repeating this process until they are satisfied with the completeness of their list of terms given their area of interest.

**Figure 6: The Tool's Topic Builder**

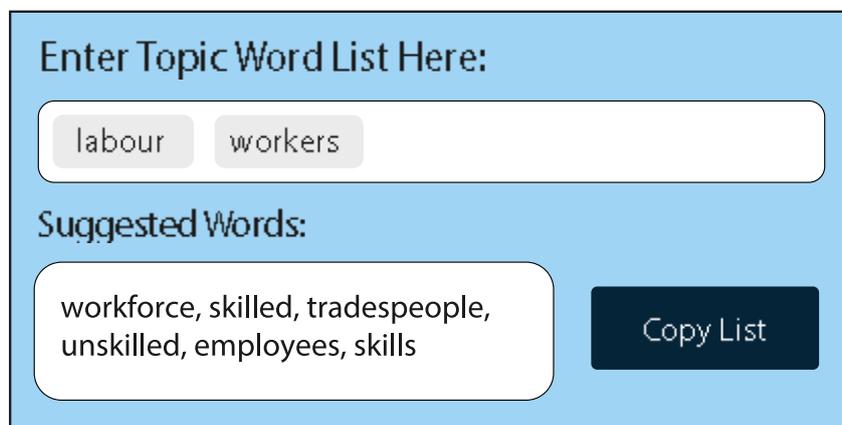

Word2Vec uses a shallow, two-layer neural network to learn word embeddings (that is, their meanings) from large text corpora. These embeddings capture semantic relationships between words, meaning that words used in similar contexts are mapped to nearby points in the vector space. We trained Word2vec from scratch on the full liaison text dataset. This allows the model to better capture the semantics of liaison notes than a model trained on a more generic corpus (such as Wikipedia or the news).[15]

---

15  Training the model specifically for our corpus was feasible because Word2Vec is a shallow neural network and requires much less computational resources or data to train than transformer-based LMs.

In this paper, our keyword-based topic and tone classifiers for "wages" and "labour costs" (used in the proceeding sections) are constructed using this approach. Lists of words about wages and labour costs were compiled in consultation with economists from the liaison team and these lists were then augmented using the Word2Vec-based topic builder.

Having defined a topic in this way, users can then use the tool to count the number of times words from their list appears in each liaison summary, assuming the frequency of mentions reflects the importance of the topic. Specifically, letting $\mathbf{d}$ be the pre-defined dictionary (list of keywords) and breaking down liaison summary $i$ into its constituent words $W_{it} = [w_1, w_2 ..., w_n]$ (with repetitions included as separate elements), we can count terms according to the below:

$$\sum_{w=1}^{W_{it}} 1\{w \in \mathbf{d}\}$$

where the sum iterates over each word, $w$, in liaison summary $i$ and the indicator function $1\{.\}$ checks if the word $w$ is in the dictionary $\mathbf{d}$. If true, it equals 1; otherwise, it equals 0. This expression is used as the numerator of Equation (1) when measured using a keyword-based approach, with the denominator equal to the total number of words in the list $W_{it}$, such that the keyword-based topic exposure indicator corresponds to the share of words in each liaison that are in the relevant dictionary.

Next, to measure the tone of each sentence, users can develop new lists of tonal keywords or rely on several pre-existing dictionaries, such as the popular Loughran and Macdonald (2016) dictionary. To create keyword topic-specific tone indices, users can first filter for paragraphs that contain relevant keywords for their topic and then calculate the tone of these pre-filtered paragraphs by counting the number of words with positive tone (as defined by the dictionary) and subtracting the number of terms with negative tone (as defined by the dictionary).

To provide an example, Table 1 shows four sentences related to wages alongside the LM-based topic probability (from BART-large-MNLI), the LM-based sentiment score (from FinBERT) and the relevant sentiment words.[16]

---

16  FinBERT classifies "higher wages growth" as having positive sentiment. This is because FinBERT was fine-tuned on labelled sentences from financial news about listed companies, where annotators were asked to give labels according to how they think the information in the sentence might affect the mentioned company's stock price. In doing so, the LM has learnt that "increasing wages" usually carries positive sentiment from the perspective of an investor. Knowing this is a feature of the model, users should interpret this information in this context.

**Table 1: Sentiment Index Examples for the "Wages" Topic**

| Sentence | LM-based topic probability | LM-based sentiment | Sentiment words |
|---|---|---|---|
| Wages rose by 4 per cent on average across the business, above the long-term average of 2-3 per cent. | 0.99 | 0.95 | "rose","above" |
| The firm has been considering different options for remuneration given the current conditions. They have decided to allow a small increase in wages, plus performance-based bonuses. | 0.97 | 0.67 | "increase" |
| The outlook for wages growth remains closely linked to CPI. | 0.99 | 0.00 | |
| Annual wage growth has been very low, at around 2 per cent, and this is likely to continue. | 0.99 | −0.96 | "low" |

Source: RBA

### 4.2.3 Assessing accuracy

The benefit of the keyword-based approach is that it is completely transparent. However, the process of developing dictionaries is manual and has several limitations. First, the dictionaries are non-exhaustive, giving rise to false negatives. Second, relying on keyword matches does not capture semantic meaning, potentially giving rise to a significant number of false positives. Finally, in our applied case, dictionaries are vulnerable to changes in the language used in liaison summaries over time, which could be driven, for example, by changing editorial preferences among management and general changes in the style of internal drafting. In contrast, the LM-based approach can capture context and semantic meaning but is less transparent and could struggle to generalise from the financial text on which it was trained to business liaison summaries. How well each approach performs is a question that must be evaluated empirically.

To assess and compare the topic accuracy of our dictionary- and LM-based classifiers, we focus on the performance of our approaches in classifying liaison paragraphs into the topics "wages" or "labour". This is done by comparing 600 labelled paragraphs from each approach to human-labelled classifications, which is the consensus classification from three human experts.

We use three statistics to summarise the validity of our dictionary and LM-based topic classifications. The first is the precision of the classifications for the wages or labour topics when measured against the human consensus – that is, given the classifier tagged a paragraph as about a given topic, what is the probability the paragraph is about that topic according to human consensus. This is estimated as the number of paragraphs both the classifier and human labels agree are about the topic, or the number of true positives (TPs), divided by the total number of paragraphs classified as being about the topic, or the number of TPs plus false positives (FPs). The second is recall – that is, given that the human consensus labelled a document as belonging to the wages or labour topics, what is the chance that our classifiers do the same. This is calculated by taking the number of TPs divided by the number of human-labelled paragraphs classified as being about that topic, or TPs plus false negatives (FNs). The final metric is the F1 score, which is the harmonic mean of precision and recall.

To select our sample of paragraphs for human assessment, we pick those that are likely to be about the wages or labour topics, as identified by the LM. We do this to deliberately increase the number of these paragraphs to make sure we have enough examples to evaluate performance properly. To adjust for this increase, or up-sampling of the positive class, we create weights for each paragraph to show how representative it is of all the liaison summaries. These weights are then used to calculate our performance statistics (see Appendix A for details).

A few things stand out in the results shown in Table 2. First, at a high level, this exercise indicates that our approaches to classifying paragraphs into topics related to economics and finance – i.e. the LM- and keyword-bases approaches – are highly accurate. Overall F1 scores are high relative to external benchmarks.[17] This is because the text of the liaison summaries being classified is very clean, having been written according to a pre-defined template that generally adheres to a thematic structure and is drafted and edited to be clear and concise. In addition to this, the topics we have chosen to assess accuracy against – "wages" or "labour" – are relatively easy to identify. In a smaller spot-checking exercise presented in Appendix B, we find that topics that require more context and nuance to properly classify – such as "non-labour costs" and "financing conditions" – appear to have lower precision (see Appendix Table B1).

### Table 2: Accuracy of our Classifiers for the "Wages" or "Labour" Topics
Benchmarked against a human consensus of 600 examples

|  | Precision (TP/TP+FP) | Recall (TP/TP+FN) | F1 |
| --- | --- | --- | --- |
| Human 1 vs. human consensus | 0.99 | 0.91 | 0.95 |
| LM-based vs. human consensus | 0.96 | 0.73 | 0.83 |
| Dictionary vs. human consensus | 0.89 | 0.97 | 0.93 |
| Source: RBA | | | |

Second, even though the overall classification accuracy is high, human error still affects the task. When comparing an individual human annotator's labels to the majority consensus of human annotators, precision is nearly perfect (0.99), meaning the annotator rarely labels something as positive when it is not. However, recall is lower (0.91), indicating that the annotator misses some of the positive labels identified by the majority.

Finally, while the dictionary-based method has better overall performance, as indicated by the F1 score, the LM-based classifier is more precise. This is because the LM is more discerning when it comes to classifying relevant paragraphs (i.e. it has higher precision) but is more prone to exclude relevant paragraphs (i.e. it has lower recall). A user looking to minimise false positives would be advised to err on the side of using the LM-based classifications.

### 4.2.4   Topics and tone: an example

For illustrative purposes, Figure 7 provides an example of the extraction of a topic exposure index (from Equation (1.1) above) and topic-specific tone index (from Equation (2) above), focusing on the "wages" topic. Indices are shown using both the LM- and dictionary-based approaches. A few features are worth highlighting.

First, for the topic exposure indices and the topic-specific tone indices, both the LM-based version and the dictionary-based version are similar when aggregated to a quarterly frequency. The correlation between each of the exposure series and topic-specific tone series is around 0.75. This similarity, or convergent validity, suggests that both indices are capturing the same underlying phenomenon, even though they were derived using different approaches.

---

17  For example, the benchmark F1 score for the BART-large-MNLI classifier when assessed against the Yahoo Answers database is 53.7 (see Davison 2020).

Second, the aggregate indices from both methodologies appear to be related to significant turning points in the official measure of wages growth. For instance, the tone about wages declined after the global financial crisis as well as immediately after the outbreak of the pandemic in early 2020. Following the pandemic, interest in the "wages" topic peaked at historically high levels owing to concerns around the availability of suitable labour following the pandemic. This increase in topic exposure among liaison contacts led a strong rebound in the official statistical of measure of wages growth. Summing up, it appears our topic exposure and topic-specific tone indices are a useful narrative tool for explaining aggregate fluctuations in topics related to economics and finance from the unique perspective of firms.

**Figure 7: Tracking the Discussion of Wages**

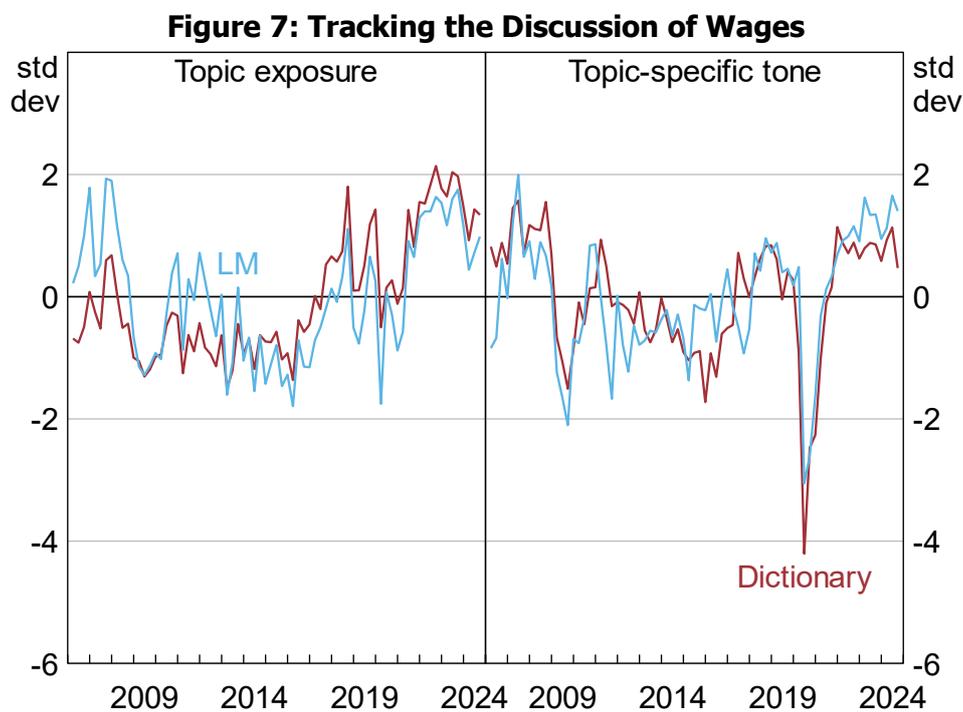

Note:    Each series is standardised to show how many standard deviations it is from its mean value.
Source:  RBA

### 4.3    Uncertainty

In addition to using keywords to capture tone, we can also use them to measure uncertainty. An important advantage of using liaison material to measure uncertainty is that it contains the direct views of firms. While there have been numerous studies focusing on measuring macroeconomic policy uncertainty (see, for example, Jurado, Ludvigson and Ng (2015), Baker, Bloom and Davis (2016), Ahir, Bloom and Furceri (2020)) there has been less focus on measuring firm-level uncertainty specifically through text-analysis. Our text analytics and retrieval tool make it possible to analyse text that records the direct views of firms over a long period of time.

The construction of our liaison-based uncertainty index is the same as that for the keyword-based topic exposure indices discussed above. As a starting point we use Loughran and McDonald's (2016) list of 298 words related to uncertainty. This initial dictionary was then consolidated by expert staff from the RBA's liaison team to a core set of words, judged to be commonly used to describe uncertainty in liaison summaries written by staff.[18] This core set was then expanded using a machine learning based model (Word2Vec) that identifies contextually similar words specific to the liaison text, leading to a final list of 83 uncertainty terms in our dictionary.

Trends in the resulting business liaison uncertainty index are shown in Figure 8. Intuitively, uncertainty peaks among liaison contacts in the global financial crisis, around the time of Australia's 2010 hung parliament, early in the pandemic and just after the first interest rate increase following the pandemic in 2022.

While our new index captures a unique aspect of uncertainty by eliciting the direct views of firms about the specific conditions they face, it is useful to externally validate it by comparing it other, more broadly based, uncertainty indices. These include the Economic Policy Uncertainty Index (EPU-Australia), the Australian Economic Uncertainty Index (AEUI) and the Australian stock market volatility index (ASX200 VIX).[19] Correlations with the EPU-Australia and AEUI are around 0.4, while the correlation with the ASX200 VIX is lower at around 0.25 (all statistically significant). This suggests that there is a common firm-driven uncertainty factor driving variation in these indices, but that our new index is not capturing much uncertainty emanating from financial markets.

**Figure 8: Business Liaison Uncertainty Index**

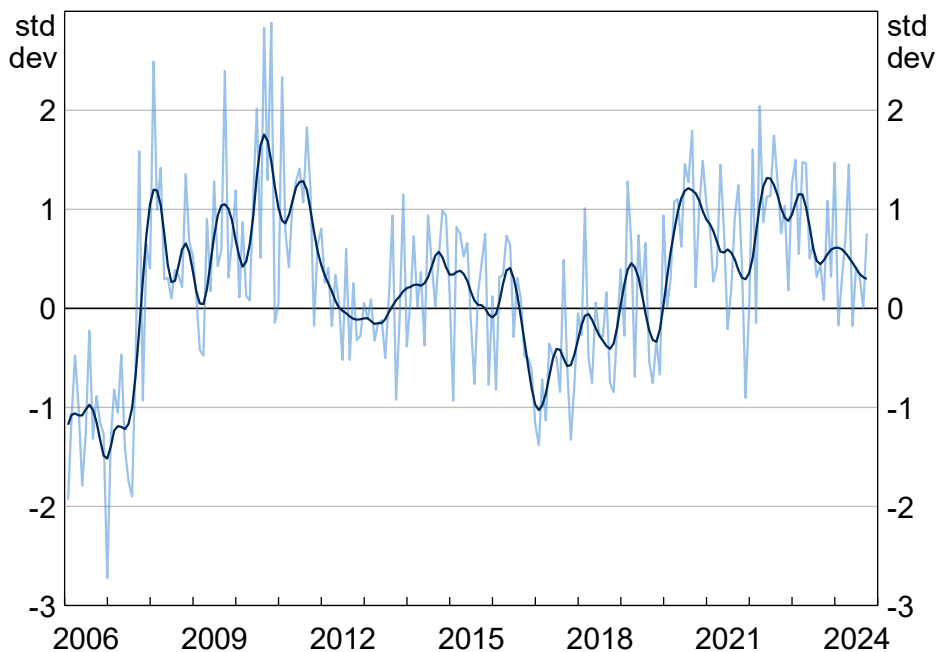

Note: Series is standardised to show how many standard deviations it is from its mean value; series is monthly with a 13-month Henderson trend.

Source: RBA

---

18  This process of expert filtering allowed staff to identify and remove words commonly used to convey uncertainty in other contexts.

19  The EPU-Australia (Baker, Bloom and Davis 2016) is a text-based frequency index counting the number of articles that are relevant to Australia that contain at least one of the words 'uncertain', 'uncertainty', 'economic' or 'economy'. The AEUI (Moore 2017), is a composite index of newspaper text, analyst earnings forecast dispersion, stock market volatility and GDP forecast dispersion.

## 4.4   Extracting precise numerical quantities

The final initial feature introduced is the ability to extract and access specific numerical quantities mentioned in the liaison summaries. The program has recorded information on firms' numerical outcomes (such as growth in their wages and prices over the previous year and expected growth for the next year) since its inception. However, except for wages, this information has not been systematically databased over the 25-year history (spanning 22,000 liaison notes) and so a time series has not been readily available for analysis. Our tool enables users to extract and access a broader range of numerical quantities. Below, we focus on the workflow associated with extracting and validating firm's self-reported inflation figures in the liaison summaries.

To create a timeseries of firms' reported price inflation, we start by extracting relevant numbers from a given firm's liaison summary. These numbers are then averaged to produce document-level price inflation rates.[20] These numbers are then averaged again across all firms, $N$, in each time period to produce quarterly (or monthly) price inflation measures, as per Equation (3) below:

$$\text{Average PriceInflation}_t = \frac{1}{N_t} \sum_{i=1}^{N_t} \text{Average inflation rate of snippets related to prices}_{i,t} \tag{3}$$

Various text processing techniques are performed in a step-by-step manner to do this:

1. First, rule-based filters are applied to narrow the search space to paragraphs that mention "price" or "prices" and that contain numbers. Paragraphs are also filtered for those that contain the words "per cent". These filters were designed in close consultation with expert staff on how liaison notes have historically been drafted and structured.

2. Next, from this narrowed corpus an LM is used to identify clauses that are likely to be about price changes. The transformer-based LM used in this stage ("roberta-base for Extractive QA") was designed to help users interrogate text by submitting questions and receiving answers about the text.[21] For each liaison sentence, the model was asked the following question: "What is the rate associated with the price change?". The model typically responds with an extract that indicates a price change, for example "4-5 per cent".

3. Next, another LM is used to determine the sign (+ or -) of the change in prices. To do this, BART-large-MNLI (the language model introduced in Section 4.2.1 to group text into pre-defined topics) is used. This model is used to determine which of the following labels best describes the sentence: "price increase", "price decrease" or "no changes to price".[22] Each label corresponds to a sign. For example, if a sentence is found to be best described by "price decrease", it will be assigned a "-1" value. A "No change" sentence is assigned a "0". The final step simply multiplies the extracted number from step two with its sign.

---

20  Because the liaison program collects information on actual and expected prices, the average includes both backward- and forward-looking price information where both are provided. A range of approaches were tried to separate past and future statements around prices (e.g. tense detection using dictionary look ups as well as a previously trained tense detection algorithm), but no method provided useful improvements.

21  Lui *et al* (2019). The Huggingface model card is here: here.

22  For each sentence, the model assigns probabilities of that sentence belonging to each of the 3 labels. These probabilities sum up to 1. The label with the highest probability value gets associated with the sentence.

4. Finally, to remove outliers, extracted prices that fall outside of the 10th and 90th percentile of the distribution across firms each year are dropped in that year.[23]

Figure 9 plots the extracted price inflation estimates against official measures of consumer price inflation as well as staff scores (on a scale from −5 to +5) for current price inflation. The series are tightly associated: the peak correlation between extracted price inflation and actual CPI inflation is 0.77 while the peak correlation with staff scores recorded for firms' prices outcome is 0.85.[24] Granger causality tests for predictive or contemporaneous relationships also indicate there is bi-directional Granger causality between the extracted self-reported price inflation series and official statistics − that is, past information in the self-reported price inflation measure can help to predict CPI inflation in the reference quarter, after considering past values of both variables, and vice versa. Moreover, the self-reported price inflation estimates lead the staff scores, with Granger causality only running in one direction.

**Figure 9: Benchmarking our Price Inflation Extractions**
Two-quarter rolling average

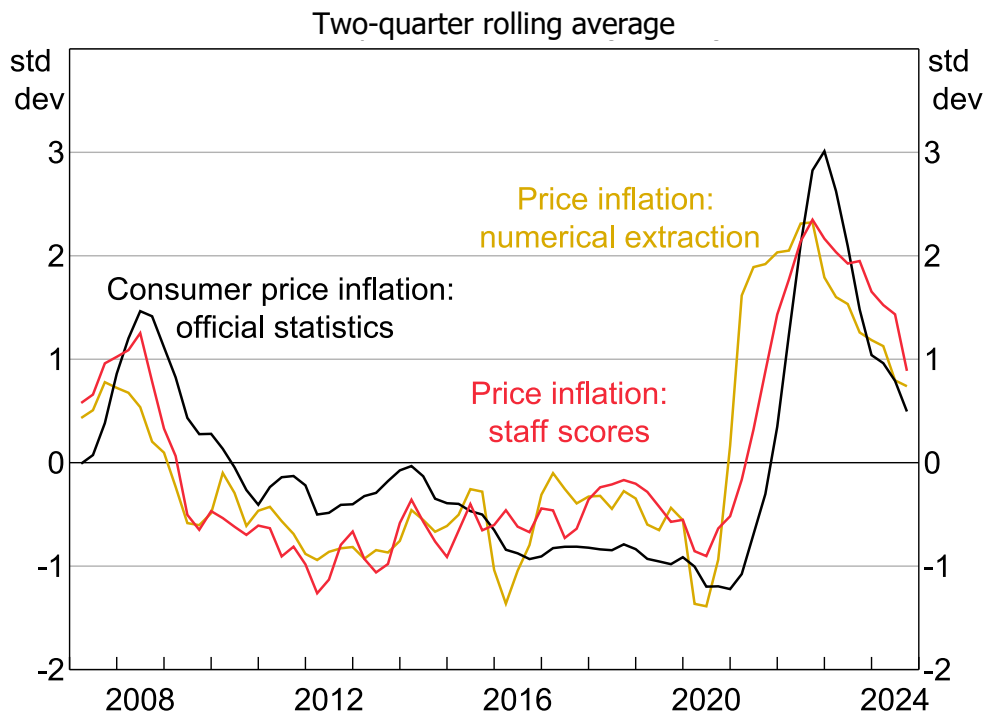

Note:     Series is standardised to show how many standard deviations it is from its mean value.
Source:   ABS; RBA

In addition to providing a timely read on developments in average price inflation over time, we can also use the new price inflation measure to examine the distribution of price outcomes across firms over time. Figure 10 illustrates how the distribution of price growth across firms has evolved. The first thing to note is that the dispersion of price inflation across firms has stayed reasonably stable over time, including over recent years, which were characterised by sizable supply shocks. However, the distribution became more positively (or right) skewed over 2021 to 2023. During that time, a few firms were reporting large price increases, while a large share of firms were reporting moderately stronger rises relative to the preceding couple of years. Relative to the decade prior to the pandemic where the distribution of outcomes was negatively (or left) skewed, firms' outcomes over 2024 remained consistent with other periods of higher inflation.

---

23  Outliers in some cases will reflect errors − either a typographical error, or extraction of a number unrelated to the firm's final price increase.

24  Because staff scores involve more judgement, it is possible the extracted series removes any human hesitancy around turning points that may affect judgement in scoring outcomes on an ordinal scale. Further, staff scores for outcomes and expectations are scored separately, whereas price extractions will often capture both and average them. As a result, the extractions draw in additional forward-looking information into the extraction measure.

**Figure 10: Distribution of Firms' Self-reported Price Inflation**
Machine-extracted estimates

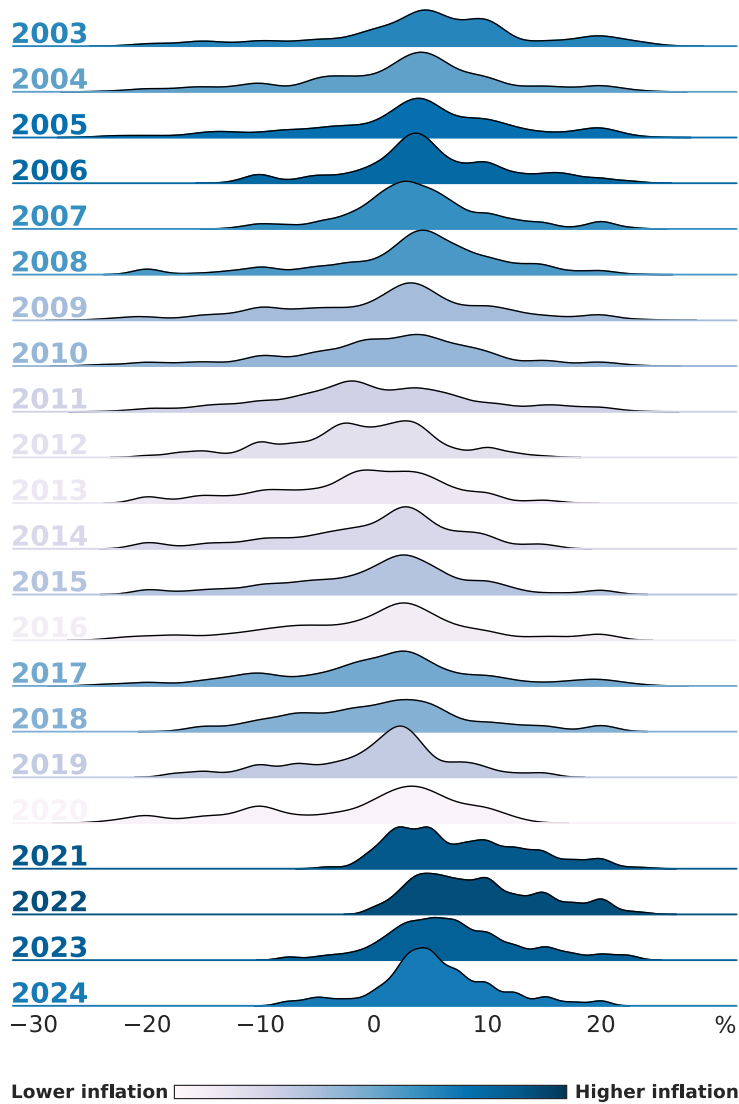

Note: When using and plotting the self-reported price inflation series we remove outliers, defined as observations falling outside of the 10th and 90th percentile of the distribution.

Source: RBA

To validate the accuracy of our approach to extracting numerical quantities, we replicate the exercise conducted above for prices on wage information. This has the advantage that the liaison team has systematically hand-recorded numerical wages growth reported by firms over the past two decades in a database, providing a benchmark to test the machine extracted numbers against.

Figure 11 plots the liaison-level errors that result from subtracting the extracted self-reported wage inflation rates from the hand-collected series. The mean and median of the errors (0.3 and 0.0 basis points, respectively) are not statistically different from zero with a mean absolute value of 50 basis points and a standard deviation of 140 basis points. Over time the two series are also very closely correlated, with a correlation of 0.92, indicating that our approach to extracting numerical quantities from the text performs well.

**Figure 11: Extracted Wage Inflation Errors**
Machine extracted vs. hand-collected

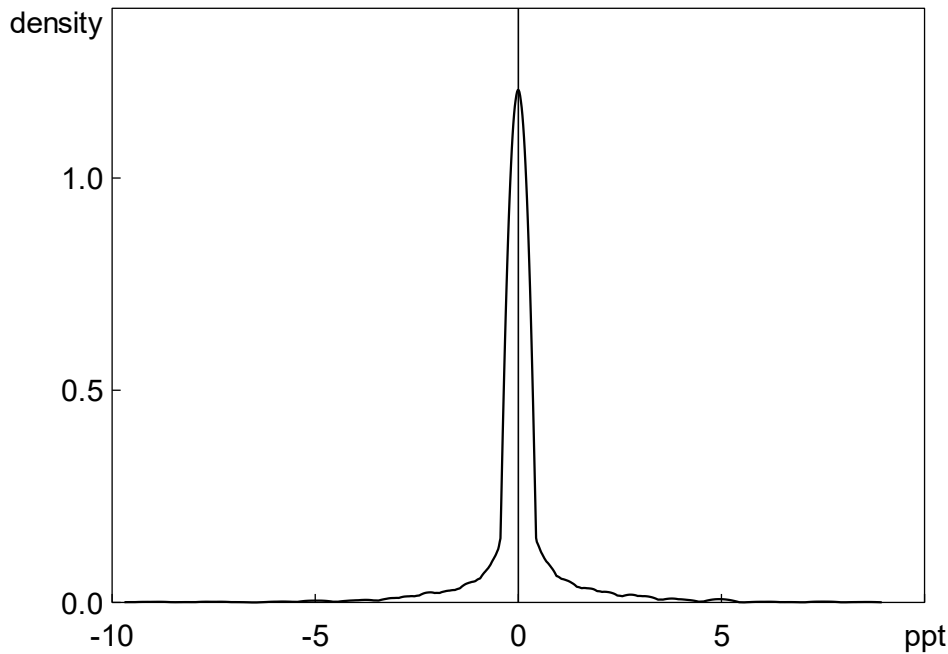

Source: RBA

## 5.    An Empirical Application: Nowcasting Wages Growth

### 5.1    Methodology

The capabilities introduced in the previous section have enabled staff to synthesise and transmit business intelligence collected through the Bank's extensive liaison program more efficiently and systematically. Among other things, use of these new capabilities over the recent period has supported the ongoing use of judgement in informing the Bank's assessment of current economic conditions – including the use of judgement informed by liaison intelligence to test and adjust forecasts derived from statistical models.

In this section, we demonstrate the potential of directly incorporating new liaison-based textual indicators into model-based nowcasts using machine learning methods. We apply this to nowcasting quarterly growth in the wage price index (WPI) for the private sector. Here, 'nowcasting' refers to estimating wages growth for the current quarter. A WPI nowcast can be an important policy input, as official statistics for the quarter are released with a 7–8-week lag.

We compare three machine learning models that incorporate all relevant text-based information from liaison to a baseline Phillips Curve model that is among the suite of best-performing models used to nowcast wages growth at the RBA. We also benchmark against the RBA's expert judgement-based adjustment to the model-based nowcast for growth in the WPI for the private sector.

We set up the exercise to be as realistic as possible.[25] In practice, the model-based nowcast for quarterly growth in the WPI for the private sector, $\Delta WPI_t$, is produced by staff around the end of the reference quarter (Figure 12). This nowcast uses the recently released figure for wages growth in the previous quarter. About one month after the model-based nowcast is produced, senior staff finalise their judgement-based nowcast for wages growth, which takes the model-based estimate and adjusts it using liaison intelligence and other timely information. This judgement-based nowcast is published in the RBA's flagship monetary policy publication, the *Statement on Monetary Policy*.

For example, imagine we were nowcasting for the June quarter 2020. Official data for the March quarter 2020, $\Delta WPI_{MQ2020}$, were released on 13 May and a model-based nowcast for the June quarter was produced shortly thereafter. One week before the publication of the *Statement on Monetary Policy* on 6 August 2020, senior staff finalised their judgement-based nowcast for wages growth for the June quarter, with the official measure, $\Delta WPI_{JQ2020}$, then published on 12 August 2020. This gives us a model- and judgement-based nowcasting error to analyse.

**Figure 12: The Nowcasting Process**

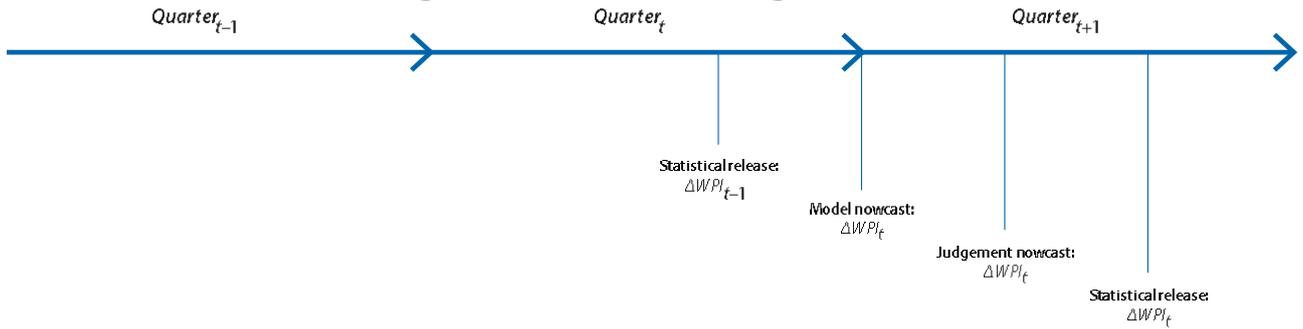

In producing the model-based nowcast staff have available to them the following baseline Phillips Curve model, which is estimated via Ordinary Least Squares (OLS) regression by minimising the sum of the squared errors:

$$\hat{\beta} = \arg\min_{\beta} \left\{ \sum_{t=1}^{T} \left( \Delta\mathsf{WPI}_t - \beta_0 - \beta_1 \Delta\mathsf{WPI}_{t-1} - \beta_2 \mathsf{UnempGap}_{t-1} - \beta_3 \mathsf{UndutilGap}_{t-1} \right. \right.$$
$$\left. \left. - \beta_4 \Delta\mathsf{InfExp}_{t-1} \right)^2 \right\}$$

Here $UnempGap$ is the gap between the unemployment rate and the Bank's latest assessment of the Non-accelerating Inflation Rate of Unemployment (NAIRU); $UnutilGap$ is the gap between the underutilisation rate and the Non-accelerating Inflation Rate of Labour Underutilisation (NAIRLU); and $\Delta InfExp$ is a measure of inflation expectations from financial markets. At the time of nowcasting, these variables are only available to a nowcaster for the previous quarter (full definitions of all variables can be found in Appendix Table C1). To estimate the gap terms, staff use a model average estimate for the NAIRU and NAIRLU; importantly, these estimates are derived from one-sided Kalman filters that only use past information.

---

25 There remains a small element of look-ahead bias in our nowcasting exercise. This is because a staff member producing the nowcast would not have had access to large pre-trained LMs over most of the sample period. Pre-trained LMs are necessary to construct several of our liaison-based textual indicators. This notwithstanding, the nowcasting improvements we report are qualitatively similar (albeit slightly weaker) if we only use our keyword-based indices (which are not subject to this look-ahead bias). Further, a nowcaster did have access to human judgement based on the liaison corpus, which was available over the entire sample period.

To this, we add 22 additional text-based variables $X_{i,t}$ and their one-period lags $X_{i,t-1}$, all extracted from the liaison corpus to create an augmented-baseline model. These additional variables include topic exposure and topic-specific tone measures for wages and labour (both LM- and dictionary-based); interaction terms between topic exposure and topic-specific tone; as well as our numerical extractions for firms' self-reported wages growth. We interact topic exposure with topic-specific tone to allow the marginal effect of tone to vary according to how much firms are talking about a specific topic – when a topic is font-of-mind, we would expect the marginal impact of a change in tone to be larger. When focusing on numerical extractions, we include the mean across all firms in each period.

Because there are over 44 covariates to estimate, we employ machine-learning-based shrinkage methods to avoid serious overfitting and associated poor nowcasting performance. The first of these is ridge regression:

$$\hat{\beta}^{\text{ridge}}, \hat{\gamma}^{\text{ridge}}, \hat{\delta}^{\text{ridge}} = \arg \min_{\beta, \gamma, \delta} \left\{ \sum_{t=1}^{T} \left( \Delta \text{WPI}_t - \beta_0 - \beta_1 \Delta \text{WPI}_{t-1} - \beta_2 \text{UnempGap}_{t-1} - \beta_3 \text{UndutilGap}_{t-1} \right. \right.$$
$$\left. \left. - \beta_4 \Delta \text{InfExp}_{t-1} - \sum_{i=1}^{22} \gamma_i X_i - \sum_{i=1}^{22} \delta_i X_{i,t-1} \right)^2 + \lambda^{\text{ridge}} \left( \sum_{j=1}^{4} \beta_j^2 + \sum_{i=1}^{22} \gamma_i^2 + \sum_{i=1}^{22} \delta_i^2 \right) \right\}$$

Ridge regression shrinks the regression coefficients by imposing a penalty on their size. The ridge coefficients minimise a penalised sum of the squared errors, with $\lambda^{ridge} \geq 0$ controlling the amount of shrinkage. Shrinkage via ridge regression is potentially well-suited to our nowcasting exercise because we have several variables (or features) that measure the same underlying concept (e.g. topic exposure) and that are highly correlated. Ridge regression shrinks coefficients that are similarly sized toward zero and those of strongly correlated variables toward each other. The penalty parameter, $\lambda^{ridge}$, is adaptively chosen via cross-validation for every nowcast (discussed below).

The second method is the least absolute shrinkage and selection operator (lasso), which is a shrinkage method like ridge, but the penalty is based on the absolute size of the coefficients rather than their square:

$$\hat{\beta}^{\text{lasso}}, \hat{\gamma}^{\text{lasso}}, \hat{\delta}^{\text{lasso}} = \arg \min_{\beta, \gamma, \delta} \left\{ \sum_{t=1}^{T} \left( \Delta \text{WPI}_t - \beta_0 - \beta_1 \Delta \text{WPI}_{t-1} - \beta_2 \text{UnempGap}_{t-1} - \beta_3 \text{UndutilGap}_{t-1} \right. \right.$$
$$\left. \left. - \beta_4 \Delta \text{InfExp}_{t-1} - \sum_{i=1}^{22} \gamma_i X_{i,t} - \sum_{i=1}^{22} \delta_i X_{i,t-1} \right)^2 + \lambda^{\text{lasso}} \left( \sum_{j=1}^{4} |\beta_j| + \sum_{i=1}^{22} |\gamma_i| + \sum_{i=1}^{22} |\delta_i| \right) \right\}$$

Because of the nature of the penalty, making $\lambda^{lasso}$ sufficiently large means lasso performs variable selection by pushing a subset of the coefficients to exactly zero during optimisation, effectively performing model selection.[26] As with the penalty parameter in ridge regression, it is adaptively chosen for every nowcast using cross-validation.

The lasso is a sparse modelling technique that selects a small set of explanatory variables with the highest predictive power, out of a much larger pool of regressors. On the other hand, the ridge regression is a dense-modelling technique, recognising that all possible explanatory variables might be important for prediction, although their individual impact might be small. An important contribution of this paper is to demonstrate that, when nowcasting wages growth in our exercise, the signal from text-based indicators extracted from business intelligence information *is sparse* rather than dense (see Section 5 for a discussion).

Finally, our third shrinkage method, *elastic net*, mixes the strengths of both the ridge and lasso methods, by performing variable selection like the lasso and shrinking together the coefficients of correlated features like ridge. The elastic net constraint is given below:

$$\lambda^{\text{elasticnet}} \left( \alpha \left( \sum_{j=0}^{4} |\beta_j| + \sum_{i=1}^{22} |\gamma_i| + \sum_{i=1}^{22} |\delta_i| \right) + (1-\alpha) \left( \sum_{j=1}^{4} \beta_j^2 + \sum_{i=1}^{22} \gamma_i^2 + \sum_{i=1}^{22} \delta_i^2 \right) \right)$$

where $\alpha$ determines the mix of the penalties and is chosen via cross-validation.

At the outset of our nowcasting exercise, we make a variety of decisions regarding our implementation strategy that are used across all three models, including defining the testing period, opting for either a rolling window or an expanding window approach and selecting a cross-validation procedure. These decisions are detailed in Table 3.

---

26  Lasso can be interpreted as placing a Laplace prior on the distribution of the coefficients, which peaks sharply around zero as opposed to Ridge which can be interpreted as a Gaussian prior.

| Table 3: Out-of-Sample Nowcasting Decisions | | |
|---|---|---|
| **Decision** | **Description** | **Choice** |
| Pure training period | Initial period used to estimate model coefficients. | March 2006 to December 2014; 36 observations |
| Nowcasting period | The period used to evaluate nowcasting accuracy. | March 2015 to September 2024; (39 nowcasts) |
| Window type | Whether to use a rolling or recursively updating window for model estimation. | Recursively updated |
| Benchmark model | Model against which nowcasting performance is compared. | Autoregressive Distributed Lag Phillips curve model |
| Data transformation | Transformation applied to the data. | All variables standardised to have a mean of 0 and a standard deviation of 1 for every nowcast |
| Evaluation metrics | Metrics used to evaluate the forecast performance. | Root Mean Squared Error (RMSE) |
| Hyperparameter range | The range of $\alpha$ and $\lambda$ values to search over when selecting an optimal model. | $\lambda$ = [0,7] 101 steps<br>$\alpha$ = [0,1] 10 steps |

We produce nowcasts for each quarter over the period March 2015 to September 2024 (a total of 39 quarters). The data available to train the model for each nowcast expands as we move forward. The first out-of-sample (OOS) nowcast, for March 2015, is based on the 36 quarters of pure training data. The next nowcast, for June 2015 uses the 36 training data points in addition to March 2015 and so on (Figure 13, top panel).

To run our ridge, lasso and elastic net regressions we must choose optimal values for the regularisation strength ($\lambda^{lasso}, \lambda^{ridge}, \lambda^{elasticnet}$) and the mixing parameter in the elastic net ($\alpha$) – the so-called hyperparameters. We do this via time series cross-validation (CV). For each of the 39 nowcasts, we remove the data point for the final quarter (that we ultimately wish to predict) and divide the remaining observations into 10 folds (Figure 13, bottom panel). These folds are used to optimally select values for $\lambda$ and $\alpha$, where the model is trained on past observations and evaluated, using the RMSE as a metric, on the last observation within each fold. Within each fold, a grid search is used to iterate through all possible combinations of the hyperparameters. The combination of hyperparameters that result in the lowest average RMSE error across all folds is then selected.

**Figure 13: Windowing Framework for In-sample CV and Out-of-sample Prediction**

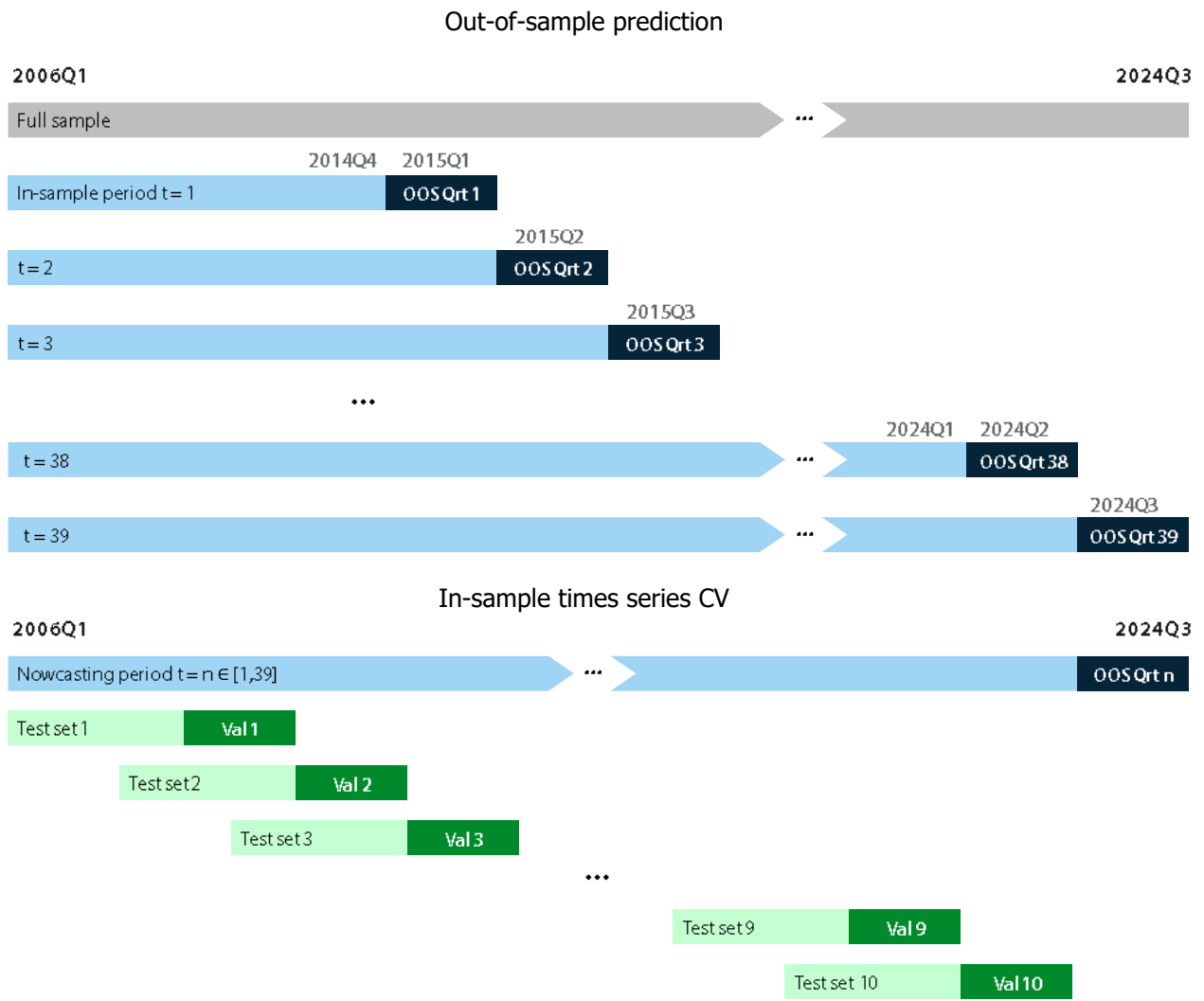

To illustrate this process, Figure 14 shows the selection of the optimal value of $\lambda^{lasso}$ and $\lambda^{elasticnet}$ (i.e. the regularisation strength) for the September 2024 nowcast. The optimal value for the ridge regression ($\lambda^{ridge}$ = 6.9) is not shown. The training data is the period from March 2015 to June 2024 and is split into the 10 CV folds. The x-axis shows a subset of the grid of values of $\lambda$ that were tested, from 0 to 0.5, and the y-axis is the average of the RMSE values from predicting the final holdout sample across all 10 CV folds. For each model, we select the value of $\lambda$ that minimises the average RMSE across all folds (optimal values are shown as dashed vertical lines for the lasso and elastic net models). In this case, the best performing model is lasso as the RMSE of the optimised lasso is slightly lower than the optimised elastic net model (and ridge, which is not shown).

**Figure 14: Optimal Hyperparameter Selection**
Cross-validated RMSEs for the September 2024 nowcast

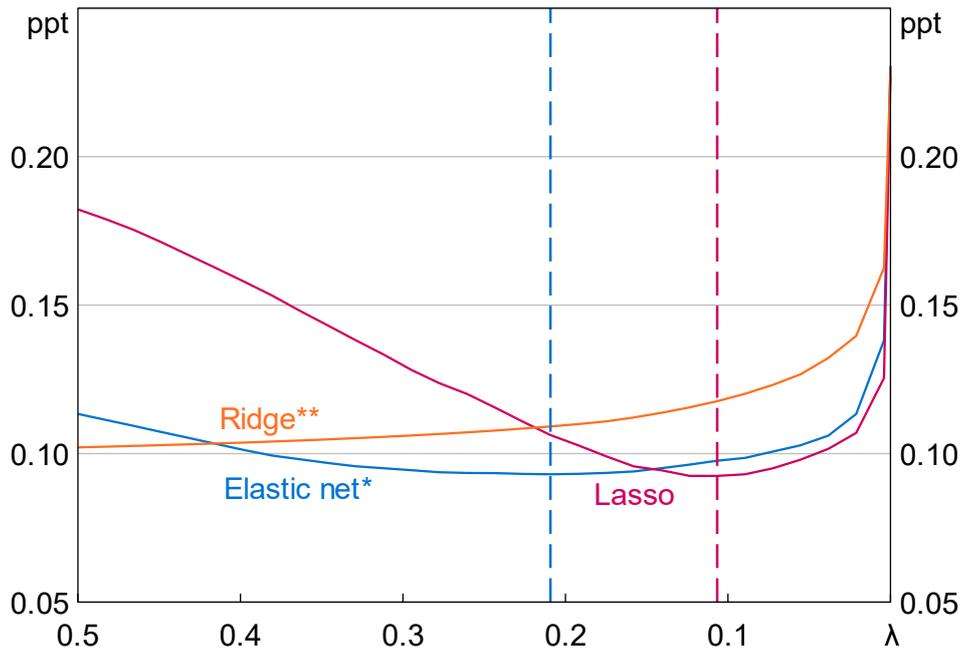

Notes:    * The mixing weight for the elastic net regression is fixed at 0.56.

           ** For illustrative purposes, the optimal value for the ridge regression ($\lambda^{ridge} = 6.9$) is not shown.

Source:   RBA

## 5.2    Results

We find that nowcasting performance significantly improves with the use of shrinkage methods and the incorporation of our text-based indicators from liaison. In the case of wages growth, at least, this underscores the ongoing use of liaison to inform the Bank's assessment of current economic conditions – including by adjusting nowcasts derived from models using judgement informed by liaison intelligence. It also highlights the usefulness of our new text analytics and information retrieval tool, which both facilitates the on-demand construction of various text-based indicators derived from the Bank's rich history of liaison intelligence and can offer a way to illustrate the relative signal of liaison information, relative to other variables, in nowcasting.

In nowcasting wages growth, the lasso model incorporating all variables achieves significant improvements of almost 40 per cent relative to the baseline Phillips Curve model estimated by OLS over the pre-COVID sample (Table 4). Most of these gains come from simply applying shrinkage methods to the current baseline Phillips Curve model. These shrinkage methods are useful in extracting a combined signal from the unemployment gap ($UnempGap$) and the underutilisation gap ($UnutilGap$). Incremental improvements are also made by incorporating the additional liaison-based indicators (plus their lags) using the lasso method. Notably, over the pre-COVID sample, the best-performing nowcast from the lasso model that incorporates all variables outperforms the expert judgement-based nowcast by 20 per cent.

Over the full sample, all shrinkage-based estimators incorporating all variables achieve significant improvements of 20 per cent relative to the baseline OLS model. Over the full sample, the incorporation of additional liaison-based information makes a larger contribution to nowcasting accuracy, improving upon the regularised-baseline by 12 basis points compared to 3 basis points over the pre-COVID sample. This highlights the usefulness of incorporating timely liaison-based information during periods when the economic landscape is rapidly evolving, as was the case during and after the pandemic. This notwithstanding, the shrinkage-based nowcasts that incorporates all variables underperform the expert judgement-based nowcast. This is because the judgement-based nowcast significantly outperformed over the first half of 2020 during the onset of the pandemic – abstracting from this period, the shrinkage-based nowcasts are on par with the performance of the judgement-based nowcast.

## Table 4: Nowcasting Performance
Target: $\Delta WPI_t$; minimums highlighted

| | Pre-COVID | | Full sample | |
|---|---|---|---|---|
| | RMSE | Ratio to baseline OLS | RMSE | Ratio to baseline OLS |
| **Baseline** | | | | |
| OLS | 0.089 | 1.00 | 0.192 | 1.00 |
| **Regularised baseline** | | | | |
| Ridge | 0.058 | 0.65*** | 0.176 | 0.91 |
| Lasso | 0.061 | 0.69*** | 0.182 | 0.95 |
| Elastic net | 0.059 | 0.66*** | 0.177 | 0.92 |
| **All variables** | | | | |
| OLS | 3.09 | 34.7 | 2.24 | 11.6 |
| Ridge | 0.064 | 0.71 | 0.152 | 0.79** |
| Lasso | 0.055 | 0.62** | 0.153 | 0.80** |
| Elastic net | 0.061 | 0.68* | 0.154 | 0.80** |
| **Judgement-based** | | | | |
| RBA published[b] | 0.069 | 0.78* | 0.133 | 0.69** |

Note: (a) Includes an additional 22 liaison-based variables and their lags; ***, ** and * denote statistical significance of nowcasting performance differences at the 1, 5 and 10 per cent levels, respectively from a Diebold and Mariano test.
(b) Using the best OLS model specification at the time, plus incorporation of other information (such as liaison messages) and judgement based on current conditions.

Sources: RBA; ABS

The results from an OLS regression including the full set of liaison indicators significantly overfits the data, resulting in poor nowcasting performance relative to the baseline model. This highlights the importance of using methods to shrink or reduce the dimensionality of the feature space.

Over the pre-COVID sample, the lasso regression applied to all variables outperforms the ridge and elastic net methods. Over the full sample, all shrinkage-based methods perform similarly. For the ridge regression to perform well over the full sample, a large degree of shrinkage is imposed over the post-COVID period. Taken together, these results suggest that sparse models may produce better predictive performance in this context.

To get a sense of the degree of sparsity, for each predictor we can examine how many times it is selected in the lasso model. That is, out of the 39 nowcasts, how many times is each predictor selected in the optimal model? The results, shown in Appendix D (see Figure D1), show that only seven predictors are included in the optimal model in more than half of the nowcasts, with around 40 per cent of the predictors never selected for inclusion in the optimal model. The finding of a sparse signal is important. Previous work by Giannone, Lenza, and Primiceri (2022) across a variety of applications in macroeconomics suggests that predictive models often benefit from including a wide set of predictors rather than relying on a sparse subset. Our result indicates this is not always the case when incorporating text-based indicators from corporate intelligence into predictive models.

Given these results, we can now examine the identity of the relevant predictors. The underutilisation gap and inflation expectations variables, which are features of the baseline Phillips Curve model, are frequently selected in the all-variables lasso model (Figure 15). This is unsurprising as their inclusion is based on strong theoretical relationships. The ordinal staff score for expected wages growth is selected in all periods. Several other traditional liaison-based measures from staff scores are also selected frequently over the pre-COVID period, including for firms' expected prices and employment intentions. Turning to the new text-based indicators, we see that firms' self-reported wages growth estimates — either hand-collected ($\Delta Wages_{t-1}^{Hand\ collected}$) or extracted using LMs ($\Delta Wages_t^{LM\ num\ extract}$) — are selected in all nowcasting periods. The frequency with which firms talk about the labour market as measured by the LM-based topic exposure ($Labour_t^{LM\ exposure}$ or its lag) also adds incremental nowcasting value in almost all periods. The liaison-based uncertainty index is also useful. Finally, in the pre-COVID period, interactions between topic exposure for wages and the associated tone of the discussion ($\Delta Wages_t^{LM\ exposure} \times \Delta Wages_t^{LM\ tone}$) is selected in almost every period, indicating that the marginal impact of tone depends on how often the associated topic is mentioned.

**Figure 15: Variables Selected for Nowcasting**
Using the lasso specification

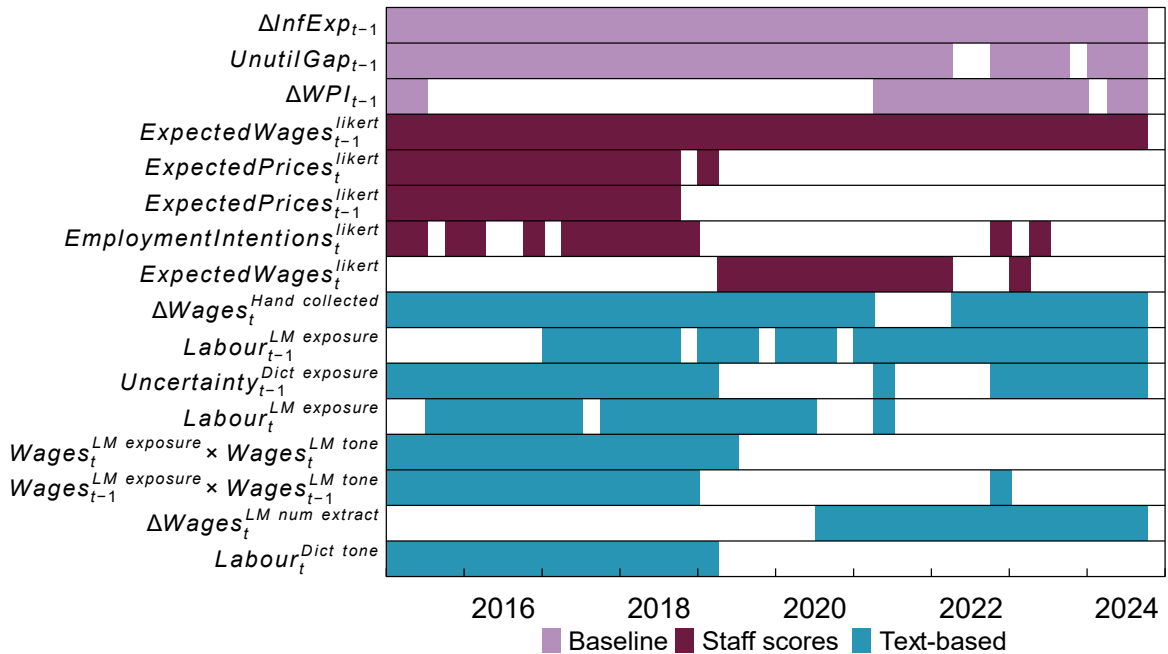

Notes:   Only showing variables that were selected in at least 10 or more nowcasts.
Source:   RBA

Summing up, integrating timely text-based liaison information into wages nowcasts makes them more responsive to new on-the-ground insights from firms and improves nowcasting performance.

Indeed, access to more timely information and increasing the responsiveness of nowcasts to incoming information were key recommendations from the RBA's internal Annual Forecasting Review conducted in November 2022 (see RBA 2022b).

To see why these lessons are important, a specific error analysis of the nowcasts is useful. Over the year to September 2022 quarterly nowcast errors for wages growth from the baseline Phillips Curve model were among the largest over our nowcasting period. In hindsight this was due to various supply-side factors that limited the available pool of skilled labour, which the baseline model does not capture very well. According to the baseline Phillips Curve model, growth in wages over the year to September 2022 was estimated to be 2.7 per cent, compared to realised growth of 3.4 per cent, with the 70 basis point error due to a series of quarterly nowcasting errors underestimating growth in wages. While liaison information was considered as part of the nowcasting process at the time, the formal incorporation of timely liaison text-based indicators would have reduced the model error to around 20 basis points. To understand why we can drill down into the micro level of the liaison text over this period. Focusing on text snippets that mention the labour market (recalling that $Labour_t^{LM\ exposure}$ was a significant predictor during this period) we can identify words or $n$-grams used in the liaison summaries over the period that that appeared at an unusually high frequency compared to other years.[27]

The results shown in Figure 16 point to a challenging labour market in 2022, characterised by staff shortages, wages pressures and the ongoing effects of the pandemic, with a significant emphasis on international factors. In practice, upward judgement informed by these liaison messages were increasingly incorporated into the nowcasts for wages over 2022, and the above results affirm that was a sensible judgement to make. However, formally incorporating this information via our text-based indicators, had the tool and these indicators been available at the time, could potentially have provided a more accurate starting point from which to apply judgement if deemed necessary.

---

27  To do this we calculate the class-based term frequency inverse document-frequency (c-TF-IDF). This is an expanded version of traditional TF-IDF, and is used in the common language embedding unsupervised topic model BerTopic for class identification. In this case our identified class is all snippets of text in 2022 that we about the "labour" topic, as identified using the LM-based approach. From this we extract terms (including bi-grams) which having the highest TF-IDF score. A high score indicates a term was used at an unusually high frequency in 2022 relative to other years.

**Figure 16: Distinctive Words Describing the "Labour" Topic in 2022**

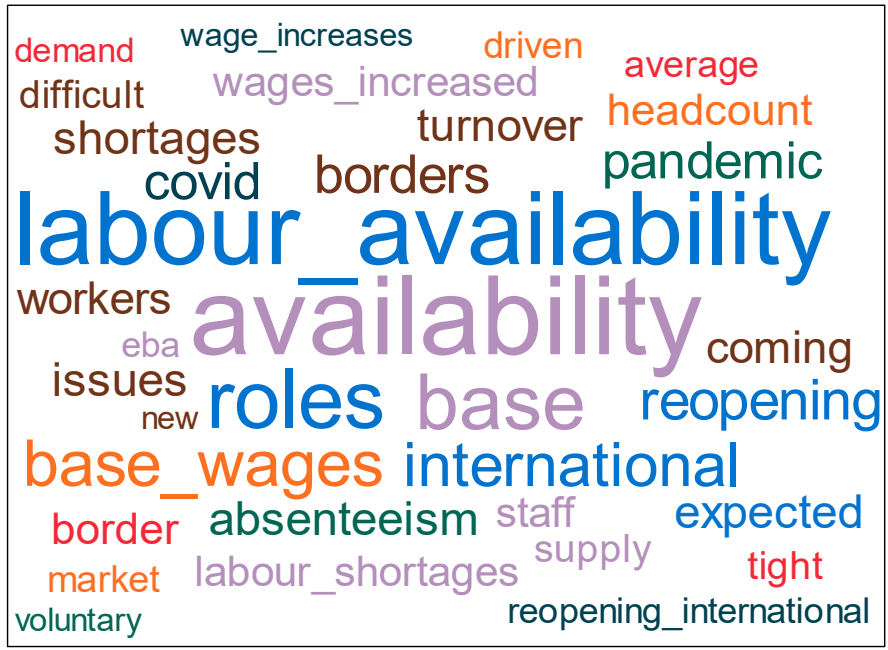

Source:   RBA

## 6.    Conclusion

Central banks rely on timely, accurate, and nuanced economic signals to assess current economic conditions and support effective monetary policy, especially during periods of heightened uncertainty. The RBA's extensive business liaison program has long provided qualitative and quantitative intelligence directly from firms, complementing insights from official statistics and other data as well as econometric models.

This paper introduces a new text analytics and information retrieval tool to enhance the analysis of the information collected through the liaison program. This system leverages recent advances in NLP to process and analyse around 25 years of confidential liaison information, covering around 22,000 firm-level interactions, with all analytics taking place within the RBA's secure information technology environment.

The RBA's new Artificial Intelligence-based text analytics and information retrieval tool enables analysts to quickly filter, query and aggregate liaison messages at scale, greatly enhancing their efficiency and ability to respond to ad-hoc demands and systematically assess information. The tool introduces several new capabilities, including rapid document querying, topic and tone classification, quantitative signal extraction (such as firms' self-reported wages growth), and uncertainty measurement from qualitative text. Our paper also presents some validation exercises showing that the new NLP-based capabilities produce accurate results when manually reviewed against human benchmarks. This notwithstanding, it will always remain critical for economists using the tool interrogate, validate and apply judgement to any LM-based outputs.

We show that integrating a large number of liaison-based indicators into model-based nowcasts for growth in the official measure of aggregate wages significantly improves nowcasting performance compared to current best-in-class baseline Phillips Curve models. Specifically, the best-performing machine learning shrinkage approach (lasso) significantly reduces nowcasting errors by meaningful magnitudes. Our results reinforce the value of liaison intelligence as well as the importance of applying shrinkage-based machine learning methods.

Overall, this work makes several key contributions. It provides the first detailed technical description of a systematised, large-scale text-based analytics and information retrieval system for central bank business liaison data. Our code is also publicly accessible to foster similar implementations and improvements across other central banks. Additionally, the paper extends an emerging literature on NLP applications in macroeconomics and highlights how capturing timely, narrative business information can materially enhance the accuracy and responsiveness of nowcasting models, especially in more uncertain economic conditions.

We consider our paper the tip of the iceberg in terms of integrating transformer-based LMs into information retrieval systems for corporate intelligence. While the LMs we work with in this paper – with 100s of millions of parameters – were considered "large language models (LLMs)" at the time of their release, today a typical LLM has billions or trillions of parameters. With more compute power, future work could use these models to improve on the capabilities we have introduced, as well as introduce new ones, such as tense detection, to construct forward and backward-looking indicators. It would also be useful to extend our empirical nowcasting exercise to other key macroeconomic variables beyond wages growth to further explore the value of central bank liaison information for predictive models.

## Appendix A: Stratified sampling to assess accuracy

Most paragraphs in the liaison corpus are not about wages (or labour costs). Figure A1 shows the distribution of the LM's predicted probabilities for the wages topic. If we set the threshold for the LM to 0.9, then it would only classify 7 per cent of paragraphs to be about wages. If we randomly sample 600 paragraphs from this population means we are unlikely to pick up many of these paragraphs – that is, we would expect to have only 42 paragraphs to assess. This can make it difficult to validate the performance of our method using metrics that rely on observing more than a few cases of wages paragraphs.

**Figure A1: Distribution of LM Topic Scores**
Share of paragraphs about "wages" by probability decile

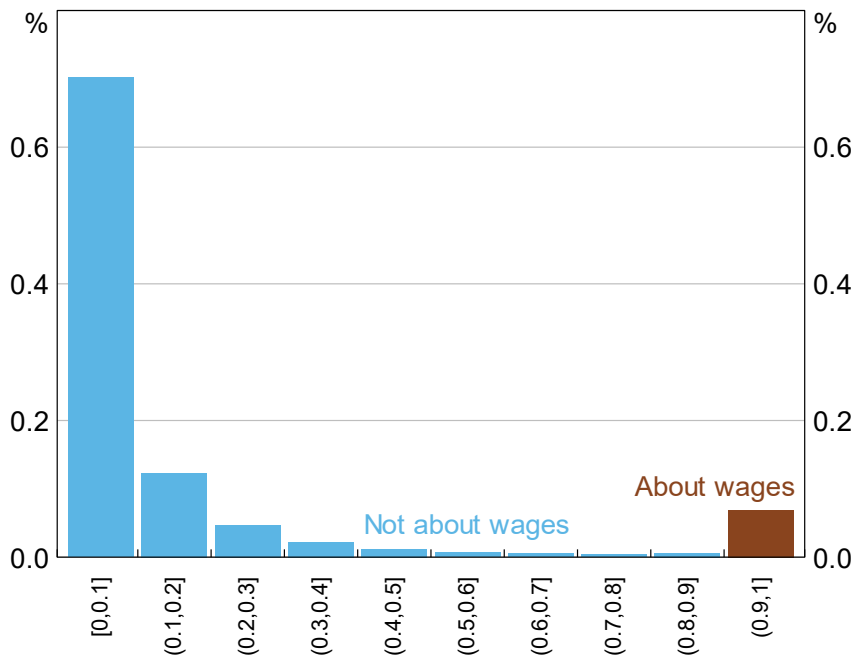

Source: RBA

The solution is to up-sample paragraphs that are likely to be about wages. We use the LM's distribution of predicted probabilities for wages to guide how we sample the population. First, we stratify the population based on the deciles of the distribution. Then we randomly sample an equal number of times from each of the deciles (or stratum). This method is known as stratified sampling.

To account for the imbalanced sampling of each stratum, we calculate an importance weight based on the sample's representativeness of the entire population. For example, a paragraph sampled from the [0,0.1] stratum (which contains about 70 per cent of paragraphs) would be 10 times more representative compared to a paragraph sampled from the (0.9,1] stratum (which is only 7 per cent of paragraphs). This is known as importance sampling and is often used to reduce the variance when estimating expected values of distributions. The importance sampling weighting formula for each of our stratums is given by:

$$w_i = \frac{N_i}{N}$$

Where $N_i$ is the number of instances in decile $i$ in the (estimated) true population, and $N$ is the total number of instances in the true population. The weighted F1 score is the sum of the F1 scores for each decile, weighted by their respective weights:

$$F1_{weighted} = \sum_{i=1}^{10} w_i \cdot F1_i$$

We can simulate a classification task on an imbalanced distribution to test if up-sampling helps improve the assessment of model performance. For the simulation, we have a population of 400,000 observations split into 10 stratums based on the distribution in Figure A1. Observations sampled from each stratum get assigned a predicted and true label based on the rates of predicted and true probabilities shown in Figure A2. The predicted probability is based on the LM's output, while the true probability is an estimation of true positive rate based on spot checks of the real liaison data. For example, predictions below 0.6 in the real data are very rarely about wages, while most paragraphs observed discussing wages have a score above 0.9.

**Figure A2: Simulation Probability Distributions**
Share of simulated draws by probability bucket

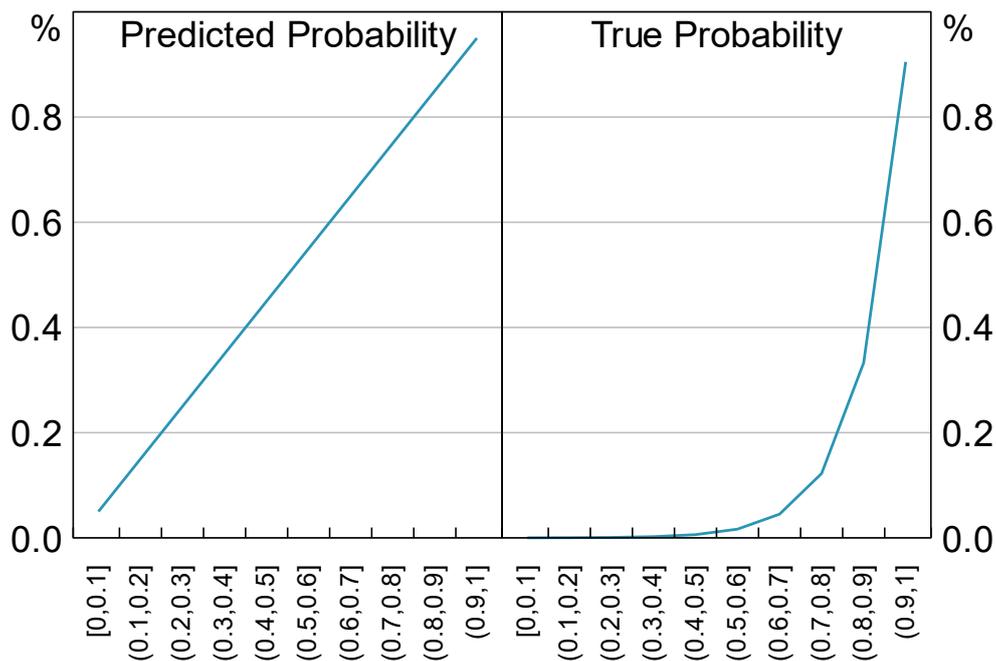

In this simulation exercise, as a baseline, we randomly sample 600 observations from the whole population and calculate our performance metrics. Then we apply stratified sampling by taking 60 random observations from all 10 stratums. This ensures we up-sample observations more likely to be true. We account for the up-sampling by including the importance weighting of each stratum when calculating the performance metrics. We repeat this 1000 times to calculate the mean and variance of each performance metric using both sampling methods.

**Table A1: Simulation Performance Metrics**
Calculated over 1000 samples

| Metric | Full Sample | Random Sampling | | Stratified Sampling | |
|---|---|---|---|---|---|
| | | *Mean* | *Variance* | *Mean* | *Variance* |
| **Recall** | 0.950 | 0.952 | 0.034 | 0.951 | 0.014 |
| **Precision** | 0.908 | 0.907 | 0.045 | 0.909 | 0.036 |
| **F1** | 0.928 | 0.928 | 0.030 | 0.929 | 0.021 |

We find that a sample of 600 observations on average gives a sufficiently accurate estimation of the true performance metrics for both methods. However, the variance of the metrics for the stratified sampling is lower than random sampling, especially for recall (almost three times smaller). This clearly demonstrates that using this stratified sampling method is more likely to result in a more accurate estimation of the performance of our model given the large class imbalanced of our data. This same method is used to up-sample wages paragraphs when validating our liaison wages measures.

## Appendix B: Spot-checking the precision of the LM-based topic classifications

In this Appendix, we perform a back-of-the-envelope assessment of the accuracy of all 14 topics classified by the transformer-based LM. To do this, we randomly sample 10 paragraphs from each topic each with an LM-based topic probability of greater than 90 per cent, for a total of 140 paragraphs. We then assess the accuracy of these LM-based classifications allowing us to do a spot check of the LM's precision for each topic. The results are presented in Figure B1 and show that topics that are more broadly defined (e.g. "costs") or open to interpretation (e.g. "financing conditions") and so requiring more context and nuance to properly classify are less precisely identified by the model.

**Table B1: Spot Checking the Precision of the LM-based Topic Classifications**
10 samples per topic

|  | Precision: TP / TP + FP (%) |
| --- | --- |
| Financing conditions | 30 |
| Costs | 50 |
| Non-labour costs | 50 |
| Prices | 60 |
| Employment | 70 |
| Margins | 80 |
| Investment or capex | 80 |
| Labour costs | 90 |
| Demand | 90 |
| Supply chains | 90 |
| Wages | 100 |
| Sales | 100 |
| Property or housing | 100 |
| Climate change | 100 |
| Overall | 78 |

Source:     RBA

# Appendix C: Variable description

## Table C1: Variables used for Nowcasting $\Delta WPI_t$

| Variable | Definition | Specification | Current[b] | Source |
|---|---|---|---|---|
| **Target variable** | | | | |
| $\Delta WPI_t$ | Growth in the wage price index for the private sector | Per cent, standardised[a] | | ABS |
| **Feature block 1: Baseline Phillips Curve features** | | | | |
| $\Delta WPI_{t-1}$ | Lag of $\Delta WPI_t$ | Per cent, standardised[a] | No | ABS |
| $UnempGap_{t-1}$ | Gap between the unemployment rate and the NAIRU (the Non-Accelerating Inflation Rate of Unemployment) | Percentage points, standardised[a] | No | ABS, RBA |
| $UnutilGap_{t-1}$ | Gap between the underutilisation rate and NAIRLU (the Non-Accelerating Inflation Rate of Labour Underutilisation) | Percentage points, standardised[a] | No | ABS, RBA |
| $\Delta InfExp_{t-1}$ | A measure of inflation expectation using an affine term-structure model | Percentage points, standardised[a] | No | RBA |
| **Feature block 2: Staff scores** | | | | |
| $Wages_t^{likert}$ | Score from −5 to +5 based on wages conditions reported by liaison contacts compared with one year ago. | Ordinal score, standardised[a] | Yes | RBA |
| $ExpectedWages_t^{likert}$ | Score from −5 to +5 based on expected wage conditions reported by liaison contacts for the year ahead. | Ordinal score, standardised[a] | Yes | RBA |
| $EmploymentIntentions_t^{likert}$ | Score from −5 to +5 based on employment intentions reported by liaison contacts for the year ahead. | Ordinal score, standardised[a] | Yes | RBA |
| $Prices_t^{likert}$ | Score from −5 to +5 based on price conditions reported by liaison contacts compared with one year ago. | Ordinal score, standardised[a] | Yes | RBA |
| $ExpectedPrices_t^{likert}$ | Score from −5 to +5 based on expected price conditions reported by liaison contacts for the year ahead. | Ordinal score, standardised[a] | Yes | RBA |
| **Feature block 3: Text-based features extracted from liaison** | | | | |
| $Wages_t^{Dict\ exposure}$ | Count of terms in the wages dictionary that appear in text snippets as a share of total words. | Per cent, standardised[a] | Yes | RBA |
| $Labour_t^{Dict\ exposure}$ | Count of terms in labour dictionary that appear in text snippets as a share of total words. | Per cent, standardised[a] | Yes | RBA |
| $Wages_t^{Dict\ tone}$ | Net balance of terms from sentiment dictionaries appearing in wages text snippets as a share of total words in wages snippets. | Index between -1 and 1, standardised[a] | Yes | RBA |
| $Labour_t^{Dict\ tone}$ | Net balance of terms from sentiment dictionaries appearing in labour text snippets as a share of total words in labour snippets. | Index between -1 and 1, standardised[a] | Yes | RBA |
| $Wages_t^{LM\ exposure}$ | Count of wages text snippets classified by an LM as a share of total text snippets. | Per cent, standardised[a] | Yes | RBA |
| $Labour_t^{LM\ exposure}$ | Count of labour text snippets classified by an LM as a share of total text snippets. | Per cent, standardised[a] | Yes | RBA |
| $Wages_t^{LM\ tone}$ | Net tone of all wages text snippets classified by an LM. | Index between -1 and 1, standardised[a] | Yes | RBA |
| $Labour_t^{LM\ tone}$ | Net tone of all labour text snippets classified by an LM. | Index between -1 and 1, standardised[a] | Yes | RBA |
| $Uncertainty_t^{Dict\ exposure}$ | Count of terms in uncertainty dictionary that appear in text snippets as a share of total words. | Per cent, standardised[a] | Yes | RBA |
| $\Delta Wages_t^{LM\ num\ extract}$ | LM-based extraction of firm-reported growth in wages compared with one year ago in liaison text. Top and bottom deciles removed from sample. | Per cent, standardised[a] | Yes | RBA |
| $\Delta Prices_t^{LM\ num\ extract}$ | LM-based extraction of firm-reported growth in prices compared with one year ago in liaison text. Top and bottom deciles removed from sample. | Per cent, standardised[a] | Yes | RBA |
| $\Delta Wages_t^{Hand\ collected}$ | Hand-collected, firm-reported growth in wages compared with one year ago. Top and bottom 15% removed from sample. | Per cent, standardised[a] | Yes | RBA |

Note:    (a) Each series over each in-sample window has a mean of 0 and a standard deviation of 1.

(b) Indicator of whether current quarter data is available at time of nowcast, otherwise only the lagged quarter is available.

## Appendix D: Variable selection in the lasso model

### Figure D1: Variables Selected for Nowcasting
Per cent of nowcasts that select the variable; lasso specification

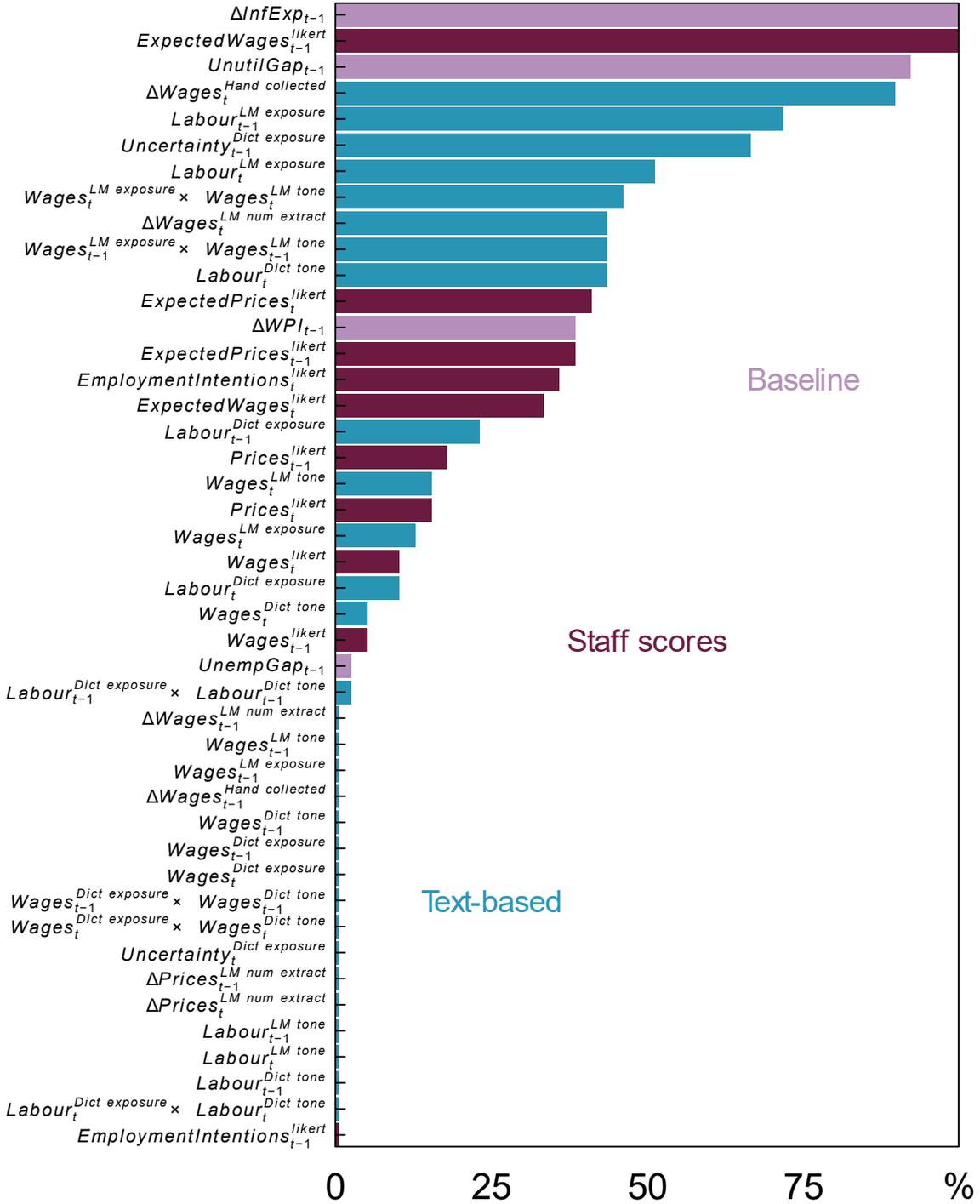